\pdfoutput=1

\documentclass[journal]{IEEEtran}
\ifCLASSINFOpdf
  \usepackage[pdftex]{graphicx}
  \DeclareGraphicsExtensions{.pdf,.jpeg,.png}
\else
\fi

\usepackage{amsmath}
\usepackage{amssymb}
\usepackage[utf8]{inputenc}
\usepackage{tabularx}
\usepackage{subfig}
\usepackage{multirow}

\begin{document}
%
\title{Buildings Detection in VHR SAR Images Using Fully Convolution Neural Networks}
%
%
%

\author{Muhammad~Shahzad,~\IEEEmembership{Member,~IEEE,}
        Michael~Maurer, Friedrich~Fraundorfer, Yuanyuan Wang,~\IEEEmembership{Member,~IEEE,} Xiao Xiang Zhu ,~\IEEEmembership{Senior Member,~IEEE,}
\thanks{M. Shahzad is with School of Electrical Engineering \& Computer Science (SEECS), National University of Sciences \& Technology (NUST), Sector H-12, 44000 Islamabad, Pakistan \& Signal Processing in Earth Observation (SiPEO), Technische Universität (TUM), Arcistrasse 21, 80333 Munich, Germany. E-mail: muhammad.shehzad@seecs.edu.pk}
\thanks{M. Maurer and F. Fraundorfer are with Institute of Computer Graphics and Vision (ICG), Graz University of Technology (TU Graz), Inffeldgasse 16, 8010 Graz, Austria. Emails: maurer@icg.tugraz.at, fraundorfer@icg.tugraz.at}
\thanks{Y. Wang and X. X. Zhu are with Signal Processing in Earth Observation (SiPEO), Technical University of Munich (TUM), Arcisstr. 21, 80333 Munich, Germany. X. X. Zhu is also with German Aerospace Center (DLR), Remote Sensing Technology Institute (IMF), Oberpfaffenhofen, 82234 Wessling, Germany. E-mails: yuanyuan.wang@dlr.de, xiao.zhu@dlr.de} 
\thanks{X. X. Zhu is the corresponding author while M. Shahzad \& M. Maurer contributed equally.}
\thanks{Manuscript revised August 2018.}}

%
%

\markboth{Submitted to IEEE Transactions on Geoscience and Remote Sensing}%
{Shell \MakeLowercase{\textit{et al.}}: Bare Demo of IEEEtran.cls for IEEE Journals}
%



\maketitle

\begin{abstract}
\textit{This is the pre-acceptance version, to read the final version please go to IEEE Transactions on Geoscience and Remote Sensing on IEEE Xplore.} This paper addresses the highly challenging problem of automatically detecting man-made structures especially buildings in very high resolution (VHR) synthetic aperture radar (SAR) images. In this context, the paper has two major contributions: Firstly, it presents a novel and generic workflow that initially classifies the spaceborne TomoSAR point clouds $ - $ generated by processing VHR SAR image stacks using advanced interferometric techniques known as SAR tomography (TomoSAR) $ - $ into buildings and non-buildings with the aid of auxiliary information (i.e., either using openly available 2-D building footprints or adopting an optical image classification scheme) and later back project the extracted building points onto the SAR imaging coordinates to produce automatic large-scale benchmark labelled (buildings/non-buildings) SAR datasets. Secondly, these labelled datasets (i.e., building masks) have been utilized to construct and train the state-of-the-art deep Fully Convolution Neural Networks with an additional Conditional Random Field represented as a Recurrent Neural Network to detect building regions in a single VHR SAR image. Such a cascaded formation has been successfully employed in computer vision and remote sensing fields for optical image classification but, to our knowledge, has not been applied to SAR images. The results of the building detection are illustrated and validated over a TerraSAR-X VHR spotlight SAR image covering approximately 39 km$ ^2 $ $ - $ almost the whole city of Berlin $ - $ with mean pixel accuracies of around 93.84\%.
\end{abstract}


%
\IEEEpeerreviewmaketitle

\section{Introduction}
%
%
%
%
\IEEEPARstart{A}{utomatic} detection of man-made objects in particular buildings from a single very high resolution (VHR) SAR image is of great practical significance particularly in applications having stringent temporal restrictions e.g., emergency responses. However, owing to inherent complexity of SAR images caused by the so-called speckle effect together with radiometric distortions mainly originating due to side looking geometry, scene interpretation using SAR images is highly challenging. Particularly in urban areas, such distortions render the data to be mainly characterized by multi-bounce, layover and shadowing effects consequently giving rise to the need of automatic and robust algorithms for object detection from SAR images.

A variety of algorithms have been published in the literature that aims at the detection and reconstruction of buildings from SAR images. Typically, most of the developed approaches rely on auxiliary information e.g., the multi-sensor data provided by the optical \cite{wegner_2011} \cite{sportouche_extraction_2011} and LiDAR \cite{tao_automatic_2011} sensors, Geographic Information System (GIS) data e.g., 2-D building footprints \cite{zhu_joint_2015}, multi-dimensional data e.g., polarimetric SAR (PolSAR) \cite{zhou_polarimetric_2016}, or multi-view/multi-aspect data such as interferometric SAR (InSAR) \cite{wegner_combining_2014}. These approaches improved the feature extraction process by providing the complimentary information. To our knowledge, the literature using only single SAR image in the context of building detection is rather sparse. Among few existing approaches, Quartulli and Datcu \cite{quartulli_stochastic_2004} employed an automatic stochastic algorithm to reconstruct buildings from a single SAR intensity image by modeling strong signals originated via dihedral scattering at the bottom and the layover at the roof edges of the building. Zhao et al. \cite{zhao_building_2013} proposed a building detection method based on marker controlled watershed algorithm. A similar approach that exploited layover and double bounce echoes to detect and determine the number of buildings from a single high resolution image was provided in \cite{cao_detecting_2014}. Ferro et al. \cite{ferro_automatic_2013} also developed a method that was primarily based on extracting a set of low-level bright (lines) and dark (shadows) primitives. Chen et al. \cite{chen_automatic_2015} introduced a more recent 1-D range detector to determine the 2-D building footprints. The method could potentially reconstruct simple symmetrical building footprints but might fail for scenes containing more complex non-symmetrical building shapes.

All the aforementioned approaches aim to extract buildings in an unsupervised (or data-driven) manner. Some researchers have also formulated the detection problem in a classification framework to benefit from well-developed supervised learning methods typically used in computer vision \cite{deng_improved_2014} \cite{zhao_support_2001}. However, the effective utilization of such supervised learning methods has two practical limitations:

\begin{enumerate}
	\item Extraction of distinctive features is necessary for reliable object detection; and
	\item A large annotated database is required which is used for training and validation.
\end{enumerate}

To address the first point i.e., distinctive feature extraction, a number of approaches have been proposed. E.g., raw pixels of images \cite{zhao_support_2001}, magnitudes of 2-D Fourier coefficients \cite{sun_adaptive_2007}, or discrete wavelet transform \cite{li_sar_2009} etc. have been used as features. Typically, feature extraction methods rely on heuristics in selecting appropriate features and therefore to cope with unaccounted situations (e.g., tolerance to incomplete views/poses in training data or randomness in speckle for different observations), expert knowledge is required to translate such discrepancies in the model for feature representation \cite{ding_convolutional_2016}.

Recently, Convolution Neural Networks (CNNs), a type of multi-layered neural networks, have significantly outperformed previous methods and became state-of-the-art in image classification. Their power lies in the fact that they directly extract \textit{high-level abstract} image features which allow replacing hand-crafted features by the machine learned features fitting to the task at hand. They have special characteristics (i.e., shared weights architecture, local receptive fields, pooling and spatial sub-sampling) that make them tolerant to high degree of image translations, skewing, scaling, rotation and other forms of geometric distortions.

\subsection{Related Work}
There exist abundant literature that employ CNNs to perform object detection in remote sensing images \cite{Vehicle2018} \cite{SAROptical2018} \cite{DeepReview2017}. In this context, we refer the interested reader to an excellent recently published survey article containing comprehensive review of deep learning techniques applied to optical remote sensing images \cite{DeepReview2017}. In contrast, the use of CNNs over SAR images is up to now limited but consistently increasing. For instance, Profeta et al. \cite{zelnio_convolutional_2016} experimented with various CNN architectures on the moving and stationary target (MSTAR) SAR dataset to achieve high classification accuracy. MSTAR dataset has also been utilized to perform SAR image segmentation in \cite{li_multiscale_2016} and \cite{malmgren_hansen_convolutional_2015}. Ding et al. \cite{ding_convolutional_2016} investigated the capability of deep CNNs to address issues in SAR target recognition, such as target translations, random speckle noise, and insufficient pose images in the training data. Utilization of CNNs in polarimetric SAR image classification has been demonstrated in \cite{zhou_polarimetric_2016}.
Some researchers also explored CNNs to solve the change detection problem in SAR images \cite{gong_change_2016}. 
Recently, the application of CNNs over TerraSAR-X spoltlight datastacks to classify built-up area has been demonstrated in \cite{zelnio_convolutional_2016}.
The problem is particularly challenging as the SAR images suffer from severe geometric distortions in urban areas and therefore they developed a robust multiscale CNN architecture to extract hierarchical features directly from SAR image patches. With the aim to develop benchmark SAR dataset, Zhao et al. \cite{zhao_convolutional_2016} also exploited CNNs over a TerraSAR-X spotlight data in image classification context and prepared a relatively large SAR image database containing five classes of object patches, including buildings, roads, vegetation, alongside and water area. They demonstrated that the CNNs trained with fairly large training samples significantly improves the classification accuracy. Xu et al. \cite{Xu2017} also demonstrated the use of CNNs over SAR images to extract buildings by manually preparing the training dataset and later incorporating modern regularization techniques (e.g., data augmentation, dropout and early stopping) to reduce testing errors.

\begin{figure*}[t]
	\centering
	\subfloat[]{\includegraphics[width=0.362\textwidth]{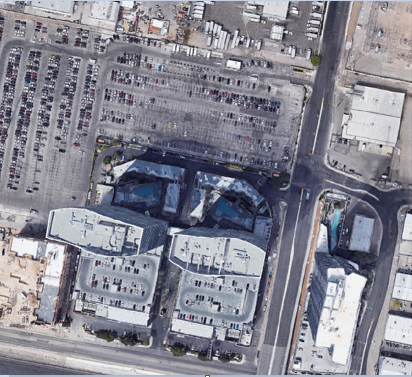}%
		\label{fig1a}}
	\hfil
	\subfloat[]{\includegraphics[width=0.39\textwidth]{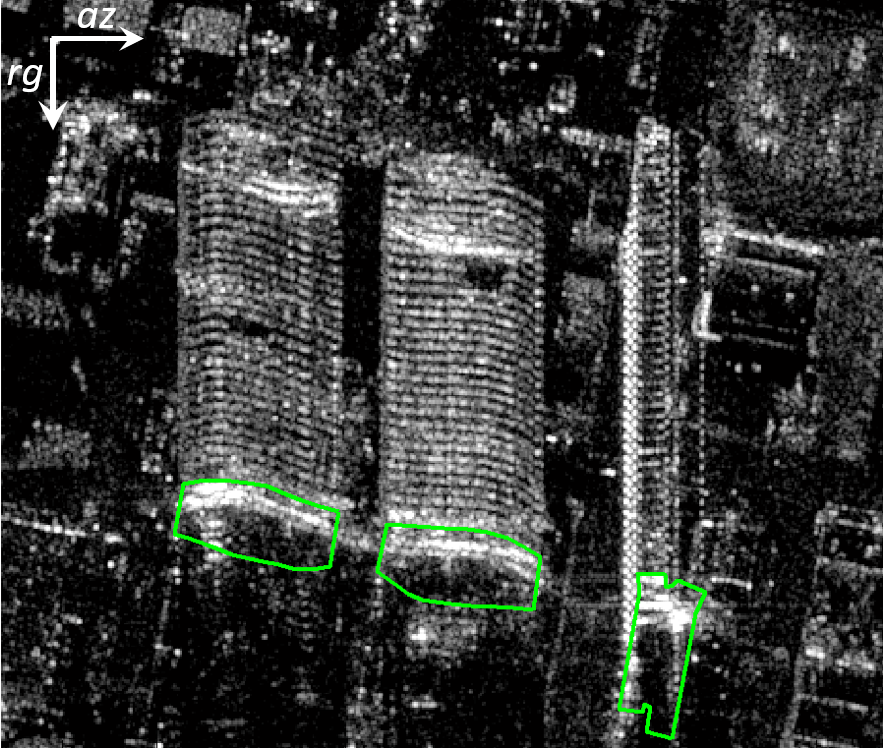}%
		\label{fig1b}}
	\caption{Depicting the challenges of SAR image interpretation together with demonstrating the limitations of directly using the \mbox{2-D} GIS building footprints onto the SAR image: (a) shows the optical image ©Google while the (b) presents the corresponding SAR image. The \textit{rg} and \textit{az} refer to the range and azimuth coordinates respectively. The three green polygons in (b) are the projections of available 2-D OSM building footprints depicted from top view in (a) onto the SAR image. It can be seen that when the illuminated scene contains elevated objects such as buildings, the so-called “layover” phenomenon (i.e., the superposition of multiple reflection sources in one pixel) occurs as a result of strong reflection of the façade in the SAR image which not only limits the direct usage of 2-D footprint projections for annotation/labeling but also makes the SAR image interpretation of urban areas highly challenging.}
	\label{fig1}
\end{figure*}

As can be imagined, the precondition for application of CNNs or any other supervised learning frameworks is the availability of annotated datasets. They are necessary not only to analyze and validate the performance of classification algorithms but are too required in the training phase where parts of annotated data are utilized to optimize prediction models. Lack of such annotated datasets is one of the major issues in application of CNNs over SAR images. Manual (or somewhat interactive) annotation, as is done in the aforementioned approaches, is one potential solution. However, due to complex multiple scattering and different microwave scattering properties of the objects appearing in the scene possessing different geometrical and material features, the manual annotation often requires expert’s knowledge (see Figure \ref{fig1}) and easily becomes impractical when large scenes need to be processed. Apart from this, another possibility of generating such a reference SAR dataset is by exploiting simulation based methods as proposed e.g., in \cite{auer_ray-tracing_2010} \cite{tao_automatic_2014} \cite{brunner_building_2010}. However such methods have their own limitations in a sense that they are either only capable of simulating simpler building shapes (e.g., \cite{brunner_building_2010}) or typically require accurate models (3-D building models and/or accurate digital surface models) to precisely generate such ground truth data which, in most cases, is not available. Thus, in view of above, \textit{automatic} annotation of SAR images, if possible, is essential.

\subsection{Significant Contributions}

The objective of this paper is twofold: First is to demonstrate the potential of automatic preparation of SAR training datasets for larger regions; Secondly, using the automatically prepared dataset to train deep CNN architecture to detect buildings in a single very high resolution SAR image. This paper extends the initial idea \cite{shahzadIGARSS2018} of automatic SAR annotation and performs a thorough analysis of the obtained SAR annotation and prediction results. Following is the novel workflow presented in this paper that involves:

\begin{enumerate}
	\item Automatic generation of annotated SAR images using spaceborne TomoSAR point clouds generated by processing SAR image stacks via advanced interferometric technique known as SAR tomography \cite{zhu_very_2010} \cite{fornaro_three-dimensional_2005} together with auxiliary information to obtain sub-image patches for training and validation;
	\item Constructing a deep Fully Convolution Neural Network (FCN) with an additional Conditional Random Field (CRF) represented as a Recurrent Neural Network (RNN) to learn a classifier via transfer learning. Such a cascaded formation has been successfully employed in computer vision and remote sensing fields for optical image classification \cite{zheng_conditional_2015} but, to our knowledge, has not been applied to SAR images;
	\item Utilizing the trained CNNs for classification of pixels as belonging to building and non-building for previously unseen input data.
\end{enumerate}

The proposed workflow leads to the following contributions to the remote sensing community. We addressed the problem of automatic generation of annotated (labelled) data which is always problematic to obtain in SAR images. In addition, we also addressed the usage of CNNs in SAR image classification which is still a relatively new research area and has not been explored much. Last but not least, since the datasets used are widely available, the annotation approach is generic and may actually lead to new perspectives in producing benchmark datasets for SAR images.

\section{Ground Truth Generation (Annotation/Labelling of SAR images)}
Annotating an image is fundamental for application of any supervised learning technique for segmentation/classification purposes. For this reason, we propose a novel workflow that utilize the SAR tomography (TomoSAR) point clouds together with auxiliary information to automatically annotate (buildings/non-buildings) SAR images of the area of interest. Before proceeding further, we briefly introduce these point clouds and later demonstrate their usage in such automatic annotation.


\subsection{TomoSAR Point Cloud}
\begin{figure}[t]
	\centering
	\includegraphics[width=3.5in]{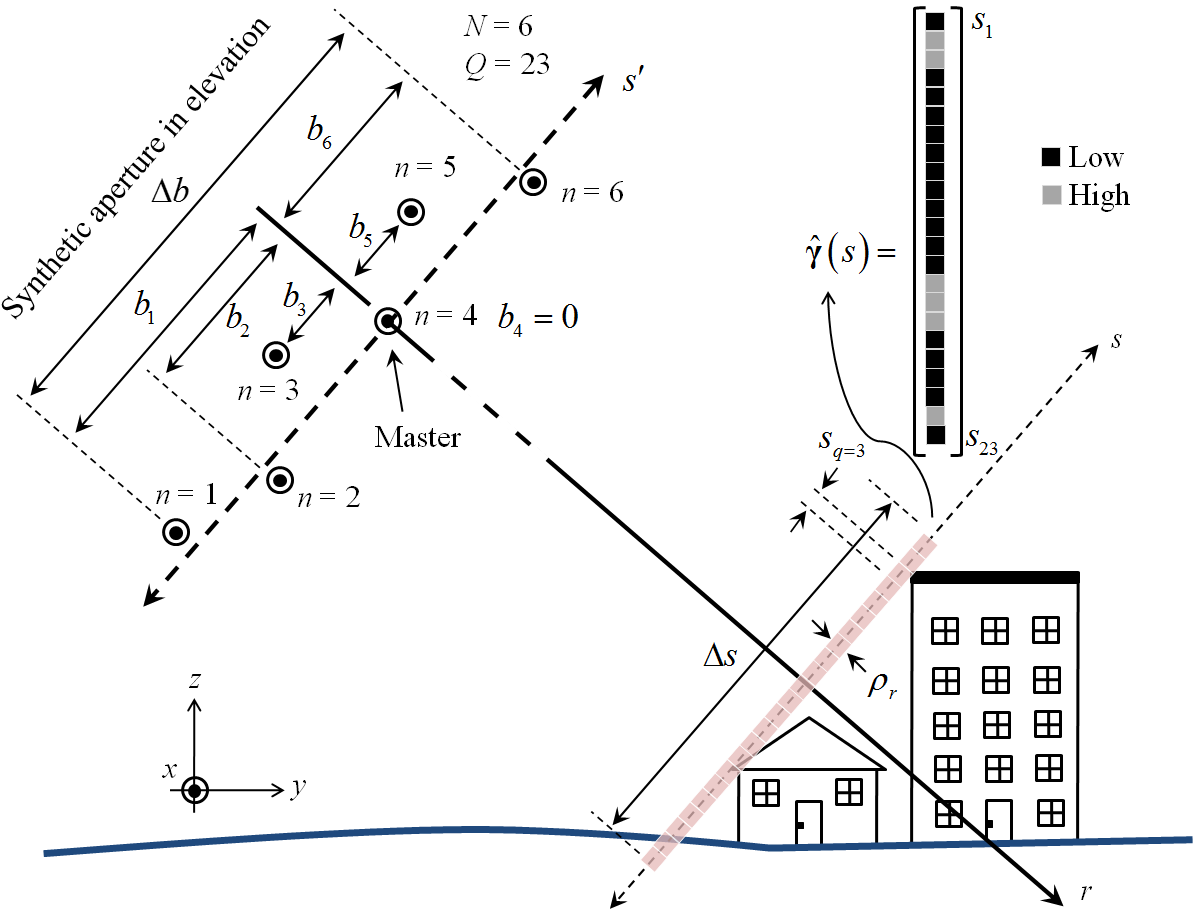}
	\caption{Schematic illustrating the TomoSAR imaging geometry. The elevation aperture is built by exploiting multi-pass/multi-baselines (six in the depicted case) from slightly different viewing angles. It is shown that the backscattering contribution from the edge of two buildings and small portion of ground are mapped onto single range-azimuth SAR image pixel. TomoSAR aims to estimate the depicted reflectivity profile $ \mathbf{\hat{\gamma }}\left( s \right) $ for discretized (pink region) elevation extent $ \Delta s $. Typically, the discretization factor is much higher i.e., $ N \ll Q $ which renders (\ref{eq:3}) to be underdetermined (i.e., infinite solutions). \textit{s} denote the elevation axis which is actually a curve but is usually approximated as a straight line due to large range distances.}
	\label{fig:TomoSAR_geometry}
\end{figure}

SAR tomography (TomoSAR) is an advanced interferometric technique that actually aims at 3-D SAR imaging. It exploits the stacked SAR images acquired from slightly varying positions to build up a synthetic aperture in the third (i.e., elevation) axis which consequently enables retrieving the precise 3-D localization of strong scatterers in a single azimuth-range SAR image pixel. The imaging geometry of SAR is shown in \mbox{Figure \ref{fig:TomoSAR_geometry}}. In the following the TomoSAR imaging model is briefly described:

Let \textit{N} represent the number of observations, the complex-valued SAR azimuth$ - $range pixel value $ g_n $ of \textit{n}th $ (n=1,...,N) $ acquisition with the corresponding perpendicular baseline $ b_n $  (see Figure \ref{fig:TomoSAR_geometry}) can be approximated as an integral of reflectivity function $ \gamma(s) $ \cite{fornaro_three-dimensional_2003} \cite{zhu_very_2010}

\begin{equation}
\label{eq:1}
{g}_{n}=\int\limits_{\Delta s}{\gamma \left( s \right)\exp \left( -j2\pi {{\xi }_{n}}s \right)}\text{d}s\text{ with }{{{\xi }_{n}}=-2{{b}_{n}}}/{\lambda r}
\end{equation}

where $ \Delta s $ denotes the span in elevation. Since it is well known that the far-field diffraction acts like a Fourier transform, the presented model is actually nothing but Fourier transform of $ \gamma(s) $ sampled at discrete frequencies (in elevation) $ \xi_n $.

The continuous model in (\ref{eq:1}) can be discretized along the elevation dimension into \textit{Q} positions (i.e., $ s_q \forall q={1,...,Q} $) by replacing the integral with the sum as follows:

\begin{equation}
\label{eq:2}
{{g}_{n}}=\sum\limits_{q=1}^{Q}{\exp \left( -j2\pi {{\xi }_{n}}{{s}_{q}} \right)\gamma \left( {{s}_{q}} \right)}+{{\varepsilon }_{n}}
\end{equation}

or alternatively in matrix form as \cite{fornaro_three-dimensional_2003} \cite{zhu_very_2010}

\begin{equation}
\label{eq:3}
\mathbf{g}=\mathbf{R}\boldsymbol\gamma + \boldsymbol{\varepsilon}
\end{equation}

where $ \mathbf{g} \in \mathbb{C}^{N \times 1} $ is the measurement vector with $ g_n \forall n \in \{1,...,N\} $, $ \mathbf{R} \in \mathbb{C}^{N \times Q} $ is an irregularly sampled Fourier transform matrix with $ {R}_{nq}=\exp \left( -j2\pi {{\xi }_{n}}{{s}_{q}} \right) $, $ \boldsymbol\gamma \in \mathbb{C}^{Q \times 1} $ is the unknown discretized reflectivity vector with $ \gamma(s_q) $, and $ \boldsymbol{\varepsilon} \in \mathbb{C}^{N \times 1} $ is additive noise usually modeled as i.i.d complex circular Gaussian random variable.

\begin{figure}[t]
	\centering
	\includegraphics[width=3.5in]{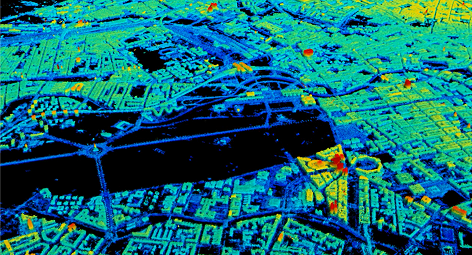}
	\caption{TomoSAR point clouds generated from TerraSAR-X data stacks of ascending and descending orbits (Site: city of Berlin). The color represents height. Black areas are temporally decorrelated objects, e.g. vegetation or water.}
	\label{fig:Berlin_TomoSAR}
\end{figure}

TomoSAR aims to invert the imaging model presented in (\ref{eq:3}) to retrieve the unknown discrete reflectivity vector $ \boldsymbol\gamma $. The reconstructed reflectivity profile along elevation axis thus allows separation of multiple layovered scatterers within single pixel \cite{zhu_very_2010} \cite{fornaro_three-dimensional_2005}. The retrieved scatterer information when geo-coded into world coordinates generates TomoSAR point clouds. Figure \ref{fig:Berlin_TomoSAR} shows the generated TomoSAR point cloud of the city of Berlin, Germany, using DLR's tomographic precessing system -- Tomo-GENESIS \cite{zhuvery2011}\cite{zhu_tomo-genesis:_2013}.

In this paper, we utilized these TomoSAR point clouds in generating labelled SAR images. The basic idea is to classify each 3-D point as belonging to buildings and non-buildings and later geo-code them back into their corresponding SAR (i.e., in azimuth and range) coordinates. The classification of each point is obtained in two ways:

\begin{enumerate}
	\item By exploiting information pertaining to already available 2-D building footprints;
	\item By classifying each TomoSAR point using an optical image classification scheme as proposed in \cite{wang_fusing_2017}. This part is not the focus of this paper. Depending on the application, a different classification technique may be employed.
\end{enumerate}

In the following, the two proposed methods to extract the building points in TomoSAR point cloud are described in detail.

\subsection{Annotation using TomoSAR point cloud and openly available OSM data}\label{sec:TomoOSM}
\begin{figure}[t]
	\centering
	\includegraphics[width=3.5in]{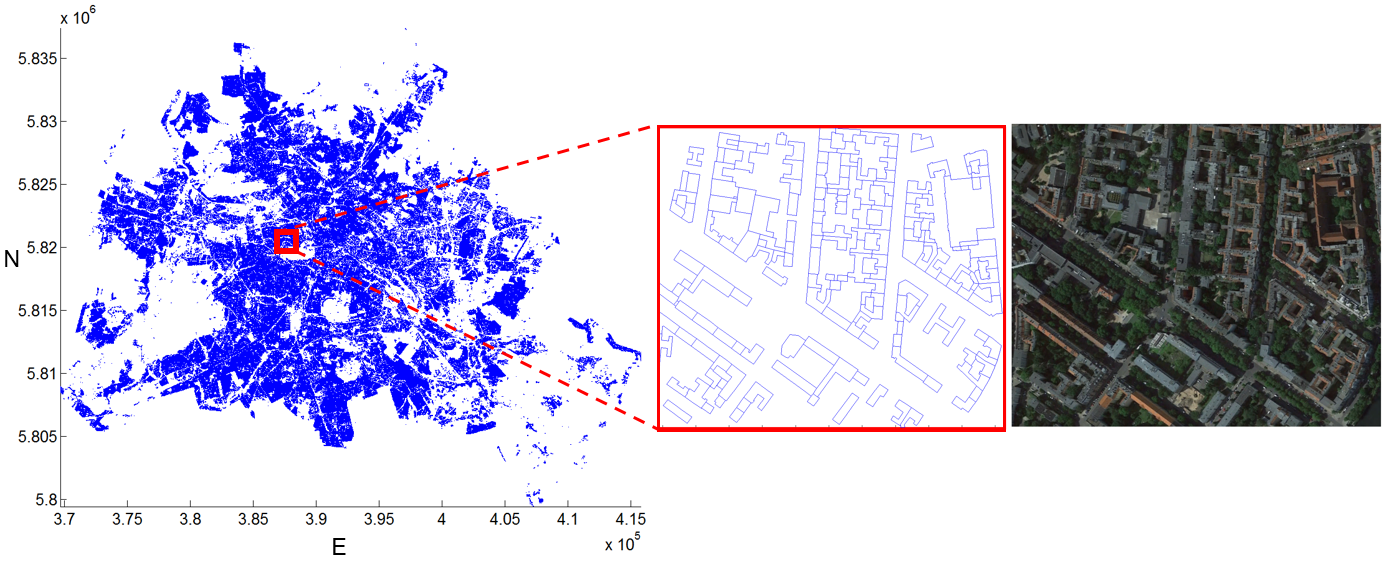}
	\caption{GIS data of Berlin from OSM: (left) 2-D building footprints; (center) Zoomed region; (right) Corresponding optical image of the zoomed region.}
	\label{fig:OSMfootprints}
\end{figure}

To classify these point clouds, the 2-D building footprints from OpenStreetMap (OSM) are downloaded from Geofabrik's website\footnote{GEOFABRIK downloads, \\ http://download.geofabrik.de/europe/germany/berlin.html} which are subsequently utilized to automatically annotate the SAR image. The OSM is based on the crowd sourcing concept and has currently around 2 million registered users \footnote{Stats - OpenStreetMap Wiki, http://wiki.openstreetmap.org/wiki/Stats} \cite{fan_quality_2014}. It consists of a large number of available building footprints with positioning accuracies varying on the order of 4m \cite{fan_quality_2014} \cite{haklay_how_2010}. The representation of building footprints is in the form of 2-D polygons having ordered list of vertices, i.e., pairs of latitude/longitude or Universal Transverse Mercator (UTM) coordinates as per WGS 84 coordinate system. The OSM data is openly available and have very high completeness percentage in many developed cities in Western Europe and US. Figure \ref{fig:OSMfootprints} shows an overview of the available 2-D building footprints in the Berlin city. The generated 3D point cloud via TomoSAR inversion using SAR image stacks is already geocoded into UTM coordinates. Now the idea is simple, we extract all those TomoSAR points that lie within the OSM building polygons. For this purpose, we employed the classical ray casting algorithm \cite{shimrat_algorithm_1962} \cite{preparata_computational_1985}. As a result, we are able to extract TomoSAR points that only belong to buildings. These building points are then projected back to SAR image coordinates (i.e., range and azimuth) to yield the building mask. 

Here one may argue that if the auxiliary information, e.g., 2-D building footprints, is being taken into account why not directly use them instead of projecting the building points in the TomoSAR point clouds back to the SAR coordinates. The rationale against this is clearly illustrated in Figure \ref{fig1}. The fact is that the inevitable side-looking SAR imaging geometry results in undesired occlusion and geometric distortions (such as layover and multi-bounce) that renders elevated objects (e.g., buildings in urban areas) to appear bright and as being projected towards the sensor consequently limiting the application potential of directly projecting such auxiliary information.

\subsection{Annotation using TomoSAR point cloud and optical image classification}\label{sec:TomoOptical}
Since the OSM is a crowd sourcing project, the lack of interest and unavailability of suitable qualified personnel especially in the underdeveloped countries may give rise to low completeness issues. Consequently, the use of OSM data for generating such reference (labeled) building masks may not be suitable. In such a case, an alternative way to extract building points in TomoSAR point cloud might be to perform semantic classification of TomoSAR point clouds. To this end, we adopt an approach \cite{wang_fusing_2017} that performs optical image classification, generates optical 3D point cloud and subsequently co-registers (fuses or matches) them with TomoSAR point clouds to achieve such labeling. Since this part is not the focus of the paper, therefore the readers are kindly referred to the original literature \cite{wang_fusing_2017} for more details. In the following, we briefly describe the main working steps of the algorithm: 

\textit{1) Optical image classification}: 
The optical images are classified patchwisely using the bag of words (BoW) method \cite{csurka_visual_2004}, which is a well-known technique in the computer vision community. Training patches are manually selected in the original image. The classification is done patchwisely in the large aerial image. The local feature used in BoW is simply the RGB value in a 3 by 3 sliding window in the patch. The classifier is a linear Support Vector Machine (SVM) \cite{cortes_support-vector_1995} implemented in an open source library VLFeat \cite{vedaldi_vlfeat:_2010}. 

\textit{2) Co-registration of optical and TomoSAR point clouds}:
An optical 3-D optical point cloud is generated from a set of nine high resolution aerial images using commercial Pix4D software \cite{noauthor_generate_nodate}. Because of the different imaging geometry of SAR and optical images, TomoSAR and optical point clouds are different in point density on façade and flat areas. This drives the co-registration algorithm to be developed in the following way:
\renewcommand{\labelenumi}{\alph{enumi}}
\begin{enumerate}
	\item \textit{Edge extraction} 
	\begin{itemize}
		\item The optical point cloud is rasterized onto a 2-D height image by computing the mean heights of points inside each 3 $ \times $ 3 grid cell.
		\item Similarly, the point density of TomoSAR point cloud is estimated on the rasterized 2-D grid by counting the number of points also inside each 3 x 3 grid cell.
		\item The edges in the optical height image and in the TomoSAR point density image are detected using Sobel filter \cite{sobel_isotropic_1968}. These edges correspond to the façade locations in the two point clouds.
	\end{itemize}
	\item \textit{Initial alignment}
	\begin{itemize}
		\item  Coarse horizontal alignment is performed by cross-correlating the two edge images while the coarse vertical alignment is achieved by cross-correlating the height histogram of the two point clouds. 
		\item These coarse alignments are fed as an initial solution to a robust iterative closest point (ICP) algorithm in the next step which provides the final co-registration solution.
	\end{itemize}
	\item \textit{Refined solution}
	\begin{itemize}
		\item The façade points in the TomoSAR point clouds are then removed, because the optical point cloud contains nearly no façade point.
		\item To refine the co-registration of the two point clouds, an anisotropic ICP (AICP) with robustly estimated covariance matrices using M-estimator is applied. Considering the large quantity of points compared with the few coregistration parameters to be estimated, the resulting coregistration accuracy is quite high \cite{wang_fusing_2017}. 
	\end{itemize}
\end{enumerate}

\textit{3) Projection of label from optical image to SAR image}: Upon successful coregistration, the 2-D classification labels from the optical images are projected to the 3-D TomoSAR point cloud using the estimated camera parameters. Each TomoSAR point classified as belonging to building is then projected to SAR coordinates. After some image morphology, a binary mask of the buildings is generated. 

\begin{figure*}[t]
	\centering
	\includegraphics[width=\textwidth]{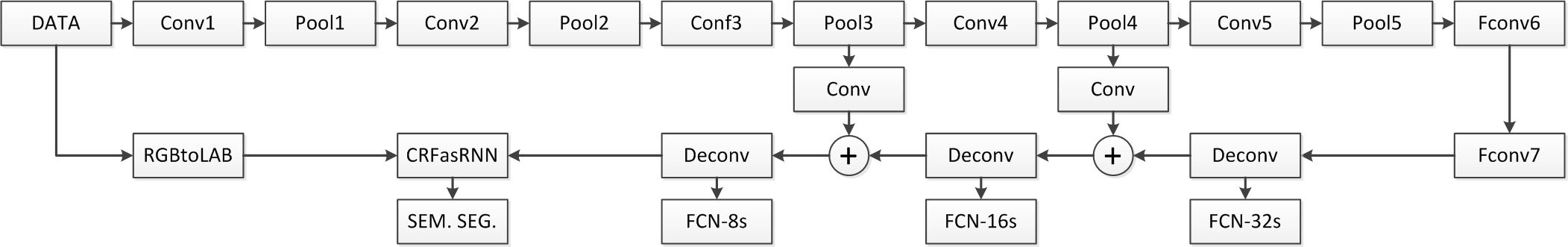}
	\caption{Overview of the semantic segmentation network. The first part of our network calculates a feature for each input pixel by exploiting a fully convolutional network (FCN) with in-network upsampling and skip-and-fuse architecture to fuse coarse, semantic, and local, appearance information. The second part of the network adds binary potentials (i.e., adding constraints to give neighboring pixels with similar intensity the same label) by using the dense CRF-RNN as proposed by \cite{zheng_conditional_2015}.}
	\label{fig:NetOverview}
\end{figure*}

\section{Architecture for SAR building detection network}\label{sec:NA}

\subsection{Brief Introduction to CNNs}\label{sec:BI}
Extracting buildings in a SAR image represents a pixel wise classification task. In computer vision this has been done using texton boost \cite{Shotton2009}, texton forests \cite{ShottonJC08} or general random forests \cite{Schulter2013}. All these methods rely on features that are hand crafted and thus prone to not always fit to the problem to classify or at least takes a lot of manual interaction to select suitable features for the specific task. Nowadays, these classification problems are tackled using Convolutional Neural Networks (CNN). One benefit of CNNs is the fact that just the structure of the network is manually designed and all the parameters, that describe how the features are calculated, are automatically learned using training data. Further, it is well known that CNNs are suitable for transfer learning. This means, that a network trained for a specific task can be reused for another task. Therefore, parts of the network can be redesigned and the unchanged part of the new network can be initialized using the parameters of the original network and fine-tuned using task specific training data. This ability of Neural Networks motivated us to use a semantic segmentation network from Computer Vision as base for our SAR Image Classification Network. Another, not negligible feature is that Neural Networks are highly parallelizable and thus suited for efficient processing using GPUs. The well-known frameworks for CNNs are Caffe or Theano. In our experiments, the Caffe framework has been employed.

\subsection{Proposed Architecture}\label{sec:PA}
The network architecture of the fully convolutional network (FCN) is based on the FCN structure of Long et al. \cite{long_fully_2015}. To additionally integrate binary potentials we add a CRF represented as RNN \cite{zheng_conditional_2015}. This gives us an end to end trainable network as depicted in Figure~\ref{fig:NetOverview}.

In detail, the first part of our network calculates a feature for each input pixel. Therefore, we exploit a fully convolutional network (FCN) with in-network upsampling and skip and fuse architecture to fuse coarse, semantic and local, appearance information \cite{shelhamer_fully_2016}. As we are using a FCN we exploit the ability to not only classify a single pixel as proposed in \cite{zhou_polarimetric_2016} \cite{li_multiscale_2016} \cite{zhao_convolutional_2016} but we perform image segmentation for input images of arbitrary size at once. Thus, we eliminate overhead calculations resulting from the sliding window approach.

The second part of the network adds binary potentials. This means it adds constraints to give neighboring pixels with similar intensity the same label. This is typically done using a Markov Random Field or to be more precise the special case of a fully connected Conditional Random Field (CRF) as presented by Krähenbühl et al. \cite{krahenbuhl_efficient_2011} whose overall energy function can be characterized as follows \cite{zheng_conditional_2015} \cite{krahenbuhl_efficient_2011}:

\begin{equation}
\label{eq:4}
E(\textbf{x}) = \sum_{i} \psi_{u} (x_i) + \sum_{i<j} \psi_{p} (x_i,x_j) 
\end{equation}

Inference of the CRF involves finding a configuration (or labeling) \textbf{x} such that the total unary $ \psi_{u} (x_i)  $ and pairwise $ \psi_{p} (x_i,x_j) $ energy components (or potentials/costs) are together minimized. Unary potentials measures the inverse likelihood (and thus, the cost) of the pixel $ i $ being assigned a label $ x_i $ while the pairwise energy components measure the simultaneous cost of assigning labels $ x_i,x_j $ to pixels $ i,j $. It typically provides an image-dependent smoothing term that favors assigning similar labels to neighboring pixels having similar properties. Specifically, in our model the unary energies are obtained from a CNN (FCN-8s architecture of \cite{long_fully_2015} as mentioned earlier). This network is primarily based on the VGG-16 network but has been modified to perform semantic segmentation instead of image classification. The pairwise energies, on the other hand, have been modeled as weighted Gaussians as follows \cite{zheng_conditional_2015} \cite{krahenbuhl_efficient_2011}:

\begin{equation}
\label{eq:5}
\psi_{p} (x_i,x_j) = \mu (x_i,x_j) \sum_{m=1}^{M} w^m G^m (\textbf{f}_i,\textbf{f}_j)
\end{equation}

where each $ G^m $ for $ m=1,2,...,M $ is Gaussian kernel applied on feature vectors. The feature vector $ \textbf{f}_i $ of pixel $ i $ is derived from image features such as RGB values and 2-D spatial location. $ w^m $ are linear combination weights while $ \mu $ is the label compatibility function which is simple Potts model $ \mu (x_i,x_j) = [x_i \neq x_j] $ in our case. 

As an end to end trainable network is preferable, we added the dense CRF represented as a Recurrent Neural Network (RNN) further called CRF-RNN as proposed by \cite{zheng_conditional_2015}.

This network was then modified to get a pixel wise two-class classification representing building and non-building.

\section{Implementation of Training Algorithm}\label{sec:ATA}
We performed staged training as mentioned in \cite{shelhamer_fully_2016} because it is less prone to divergence. First the single-stream FCN-32s is trained then the training continued with the two-stream FCN-16s and the three-stream FCN-8s. Next, the CRF-RNN is added and trained by keeping the FCN-8s part constant. Finally, a fine-tuning of the complete network has been performed. Each stage was trained for 400,000 iterations with constant learning rate ($ 1e^{-10} $, $ 1e^{-12} $, $ 1e^{-14} $, $ 1e^{-12} $ and $ 1e^{-12} $ for each stage respectively) a momentum of 0.99, weight-decay of 0.0005 and a pixel wise soft-max loss (that has been averaged over 100 images each epoch).

As the network contains convolutional layers as well as pooling layers the resulting segmented image is reduced in dimension. This is compensated by in-network upsampling layers whose parameters are initialized as bilinear filtering and further refined while training. Moreover, as suggested by \cite{zheng_conditional_2015}, in all our experiments, during training, we fixed the number of mean-field iterations in the CRF-RNN to 5 to avoid vanishing/exploding gradient problems and to reduce the training time. However, the number of iterations were raised to 10 for deploying/inference (when evaluating the test images). Moreover, the compatibility transform parameters of the CRF-RNN were initialized using the Potts model.

\section{Experimental Results \& Validation}\label{sec:ERV}

\subsection{Dataset Description}\label{sec:DS}
To validate our approach, we employed SAR datasets consisting of a TerraSAR-X high-resolution spotlight image,  and a 3-D TomoSAR point cloud of Berlin. The SAR image has spatial resolution of about 0.588m and 1.1m in range and azimuth directions respectively. The image was acquired from ascending orbit with incidence angle of 36$^{\circ}$ which almost provides a full coverage of the whole city. The 3-D TomoSAR point clouds have been generated from stacks of 102 TerraSAR-X high spotlight images from ascending and descending orbits covering almost the whole city of Berlin  using the Tomo-GENESIS software developed at the German Aerospace Center (DLR) \cite{zhuvery2011} \cite{zhu_tomo-genesis:_2013}. The number of points in the Berlin data set is approximately 30 million.

The optical images used for annotation were attained in March 014 and include nine UltraCam aerial images of Berlin having an altitude of around 4000 m. The ground spacing is roughly 20 cm per pixel. The camera positions and orientations were measured by an onboard GPS and inertial measurement unit with standard deviations of about 5 cm and 5 $ \times $ 10$^{-4 \circ}$, respectively.

\subsection{Results of Automatic Annotation}\label{sec:RAA}

\begin{figure}[t]
	\centering
	\includegraphics[width=3.5in]{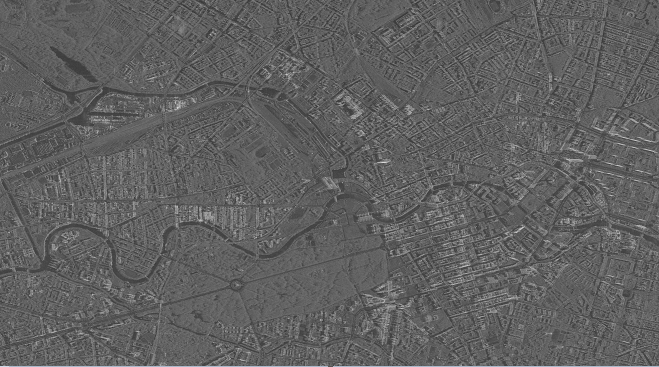}
	\caption{SAR intensity image covering almost the whole city of Berlin.}
	\label{fig:MainImage42}
\end{figure}

\begin{figure}[t]
	\centering
	\includegraphics[width=3.5in]{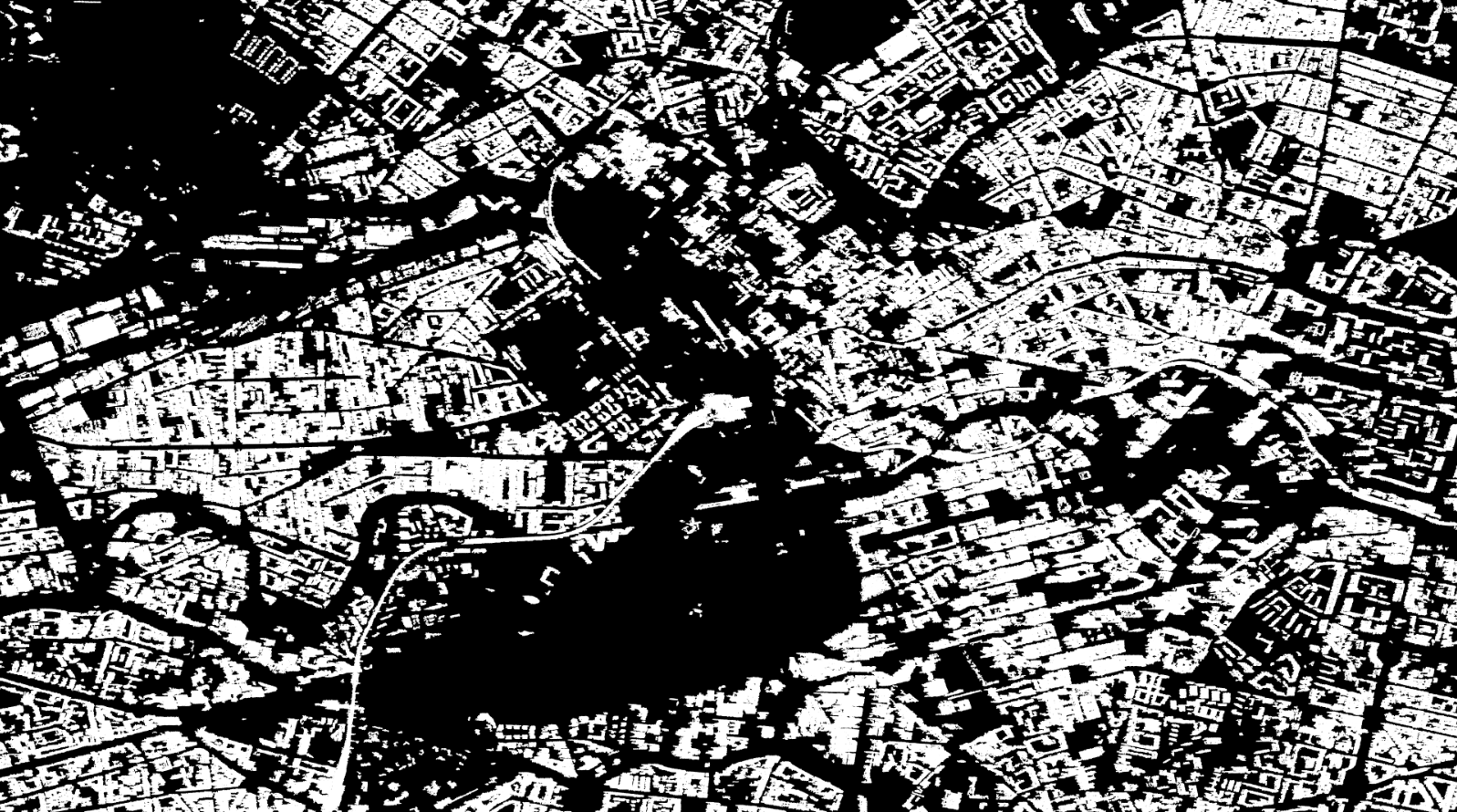}
	\caption{Automatically generated mask of building regions using OSM + TomoSAR point clouds for the SAR intensity image shown in Figure \ref{fig:MainImage42}.}
	\label{fig:MainImage42_labelled}
\end{figure}

Figure \ref{fig:MainImage42} shows the SAR intensity image covering almost the whole city of Berlin (around 39 km$ ^2 $) while Figure \ref{fig:MainImage42_labelled} demonstrates the resulting mask of building regions obtained automatically using the OSM building footprints. Since the completeness percentage of OSM data is quite high for many cities in Europe and US, it can be seen that automatic annotation/labeling using this data is quite generic and has the potential of producing benchmark SAR datasets which is still missing within the relevant community. However, although quite a lot of buildings are present, it is also worth to mention that since it is a crowd sourcing project, there are still few missing buildings and inner yards. Figure \ref{fig:Missing_Buildings} shows a couple of such examples. In addition to this, there are also false annotations like some parts of the railway tracks originating from the Berlin central station have also been labelled as building structure in the OSM data (see Figure \ref{fig:FalseRailway}). As a consequent, when OSM data is utilized to extract building points in the TomoSAR point cloud, points belonging to such railway tracks are misclassified as buildings and when projected back to SAR image coordinates yields false annotation/labelling. Although limited but on the other hand, the use of optical image classification and TomoSAR point cloud avoids this false labeling as depicted in Figure \ref{fig:FalseRailway_d} and produces better annotation results but may be restrictive in a sense to generate large-scale datasets. 

\subsection{Accuracy Analysis of Automatic SAR Annotation}
To perform the precise accuracy analysis of the produced annotations, we have manually labeled the building pixels in the SAR image covering an area of around 3.3 km$ ^2 $ in the Berlin city. Figure \ref{fig:SARImage-TrainingAnalysis} shows the selected SAR image while Figure \ref{fig:GT} shows its corresponding ground truth (GT) annotation obtained by manual labeling of building pixels/regions in the selected SAR region. For qualitative evaluation, Figure \ref{fig:CommonOSMOpt} and \ref{fig:DiffOSMOpt} shows the common and difference maps for visual comparison. The difference maps are obtained by subtracting the produced annotation masks OSM-Ref and Opt-Ref from the GT annotated mask respectively. The green pixels in Figure \ref{fig:DiffOSMOpt} indicate no change while the red pixels denote the missing buildings and the blue pixels shows the regions labelled as buildings in the generated building masks using the two proposed annotation schemes but not present in the GT reference mask. For quantitative evaluation, Table \ref{table:Accuracy_Annotations} shows the performance of the proposed annotation schemes using the common and difference maps by employing the standard precision/recall evaluation metrics computed as $ \text{Precision (\%)} = 100 \times (\frac{t_p}{t_p+f_p}) $ and $ \text{Recall (\%)} = 100 \times (\frac{t_p}{t_p+f_n}) $ where $ t_p $ are the number of white pixels (true positives) in the common image while $ f_n $ and $ f_p $ are the number of red (false negatives) and blue pixels (false positives) in the difference image respectively. 

The evaluation statistics in Table \ref{table:Accuracy_Annotations} depicts that both the proposed annotation methods correctly label building pixels with good accuracy. However, in terms of completeness, OSM-Ref shows less relative accuracy owing to the already mentioned fact that few buildings are missing in the crowd sourced OSM building footprint data. In this context, the use of accurate cadastral maps may help in achieving high degree of recall/completeness. 

\begin{table}[t]
	\caption{Quantitative evaluation statistics of automatically produced SAR annotated masks.} 
	\centering  
	\begin{tabularx}{2.5in}{c c c} 
		\hline\hline                        
		Evaluation Metrics & OSM-Ref & Opt-Ref \\ [0.5ex] 
		\hline                  
		$ t_p $ & 5614059 & 6580131 \\  
		$ f_p $ & 1191211 & 1290182 \\
		$ f_n $ & 1573086 & 607014  \\
		$ t_n $ & 12408130 & 11343087 \\ 
		\text{Precision (Correctness \%)} & 82.49 & 83.61 \\ 
		\text{Recall (Completeness \%)} & 78.11 & 91.55 \\ [1ex]    
		\hline  
	\end{tabularx}
	\label{table:Accuracy_Annotations} 
\end{table}

In the following, we present the experimental results and its analysis obtained by employing the deep learning based staged network architecture exploiting both these automatic annotations.

\begin{figure}[t]
	\centering
	\includegraphics[width=3.5in]{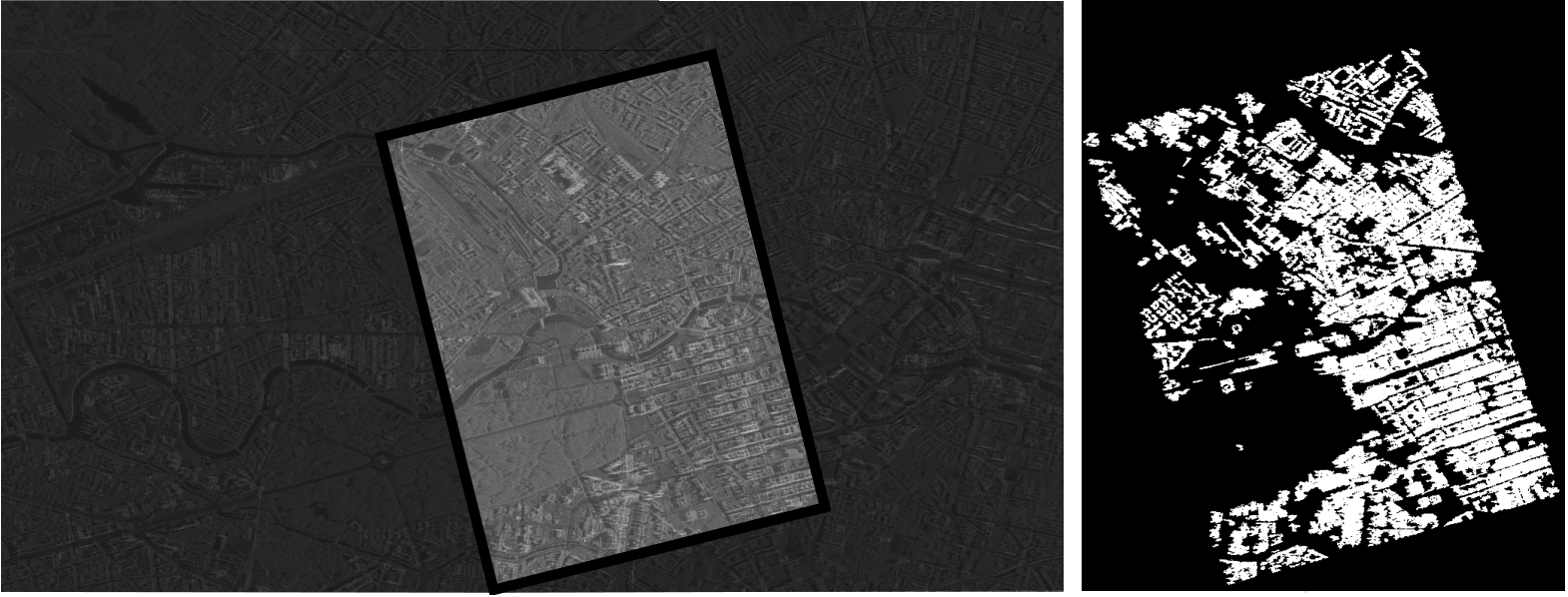}
	\caption{SAR intensity image partly covering the city of Berlin $ - $ highlighted region (left) and corresponding generated mask of the highlighted region (right) using optical image classification + TomoSAR point clouds.}
	\label{fig:MainImage42_Optical}
\end{figure}

\begin{figure}[t]
	\centering
	\includegraphics[width=3.5in]{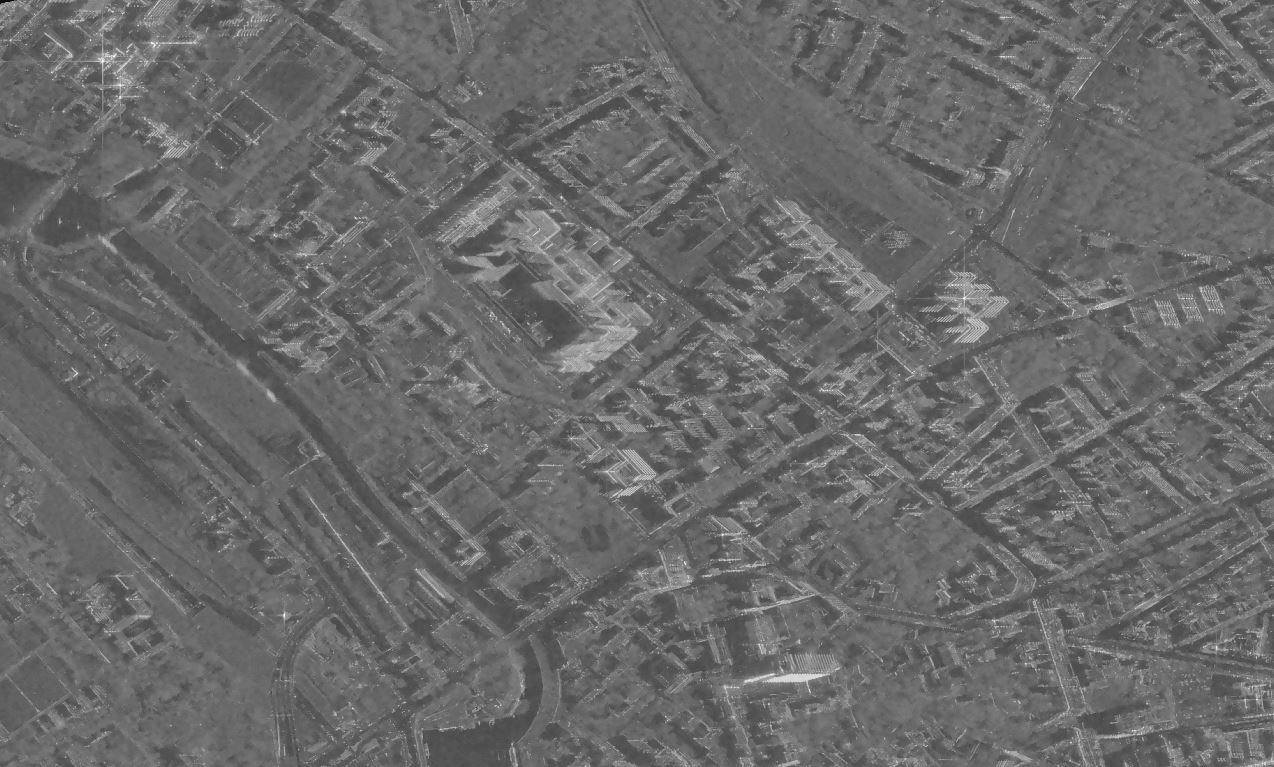}
	\caption{The SAR image of the selected 3.3 km$ ^2 $ area with following UTM coordinates: 33U top left (389072 E, 5822399 N), bottom left (388741 E, 5820939 N), top right (391201 E, 5821922 N) and bottom right (390900 E, 5820460 N).}
	\label{fig:SARImage-TrainingAnalysis}
\end{figure}

\begin{figure}[t]
	\centering
	\includegraphics[width=3.5in]{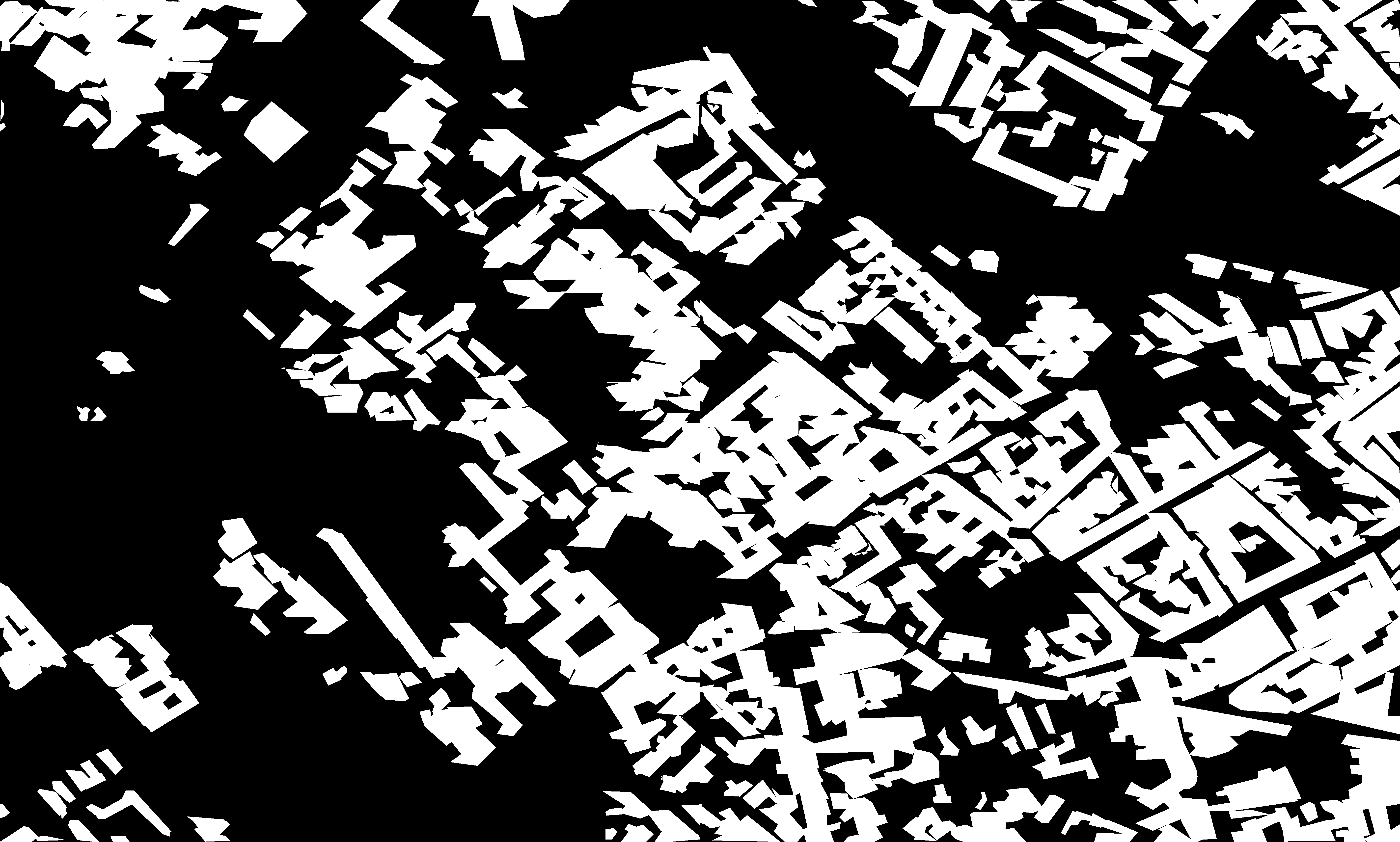}
	\caption{Ground truth (GT) mask obtained after manual labeling of building pixels/regions of the SAR image depicted in Figure \ref{fig:SARImage-TrainingAnalysis}. The mask is used for accuracy analysis of the generated SAR annotations using the two proposed schemes.}
	\label{fig:GT}
\end{figure}

\begin{figure}[t]
	\centering
	\subfloat[]{\includegraphics[width=0.5\textwidth]{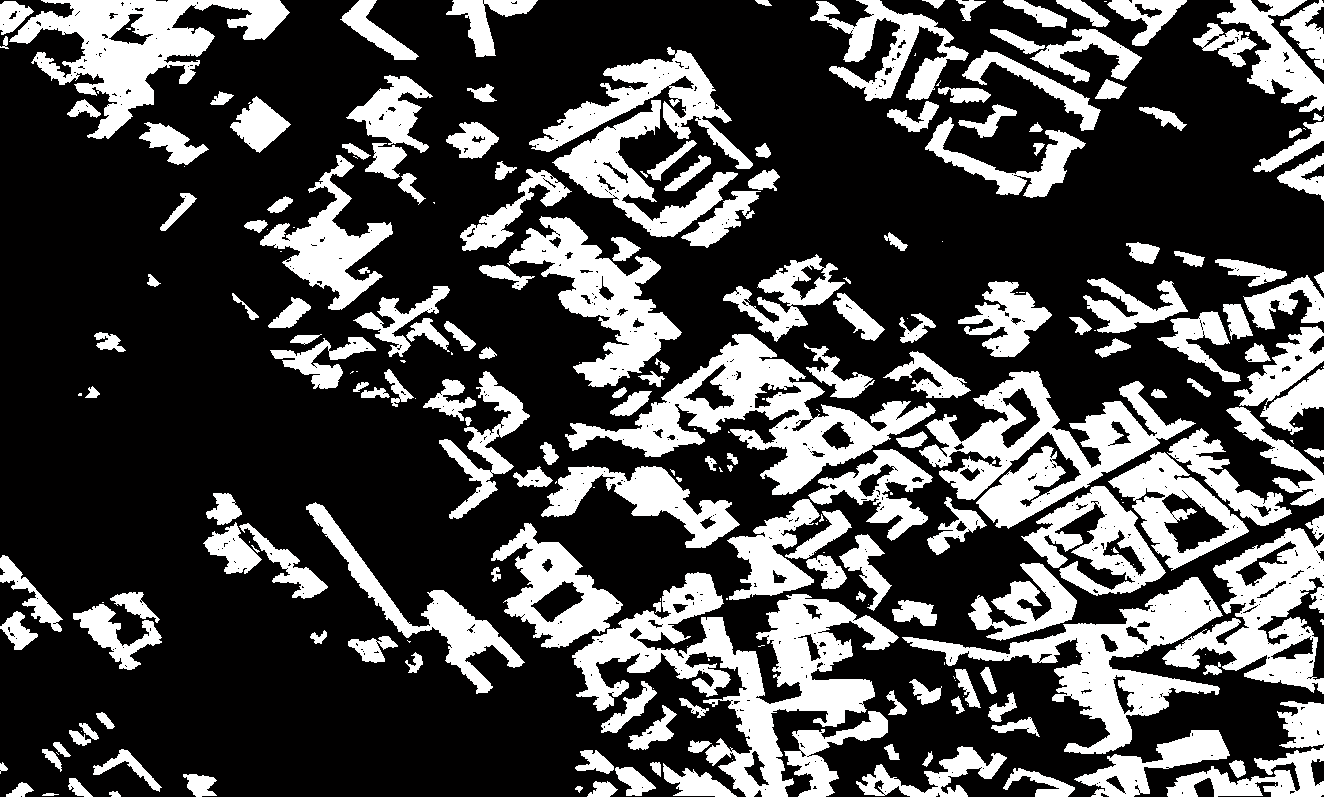}%
		}
	\hfil
	\subfloat[]{\includegraphics[width=0.5\textwidth]{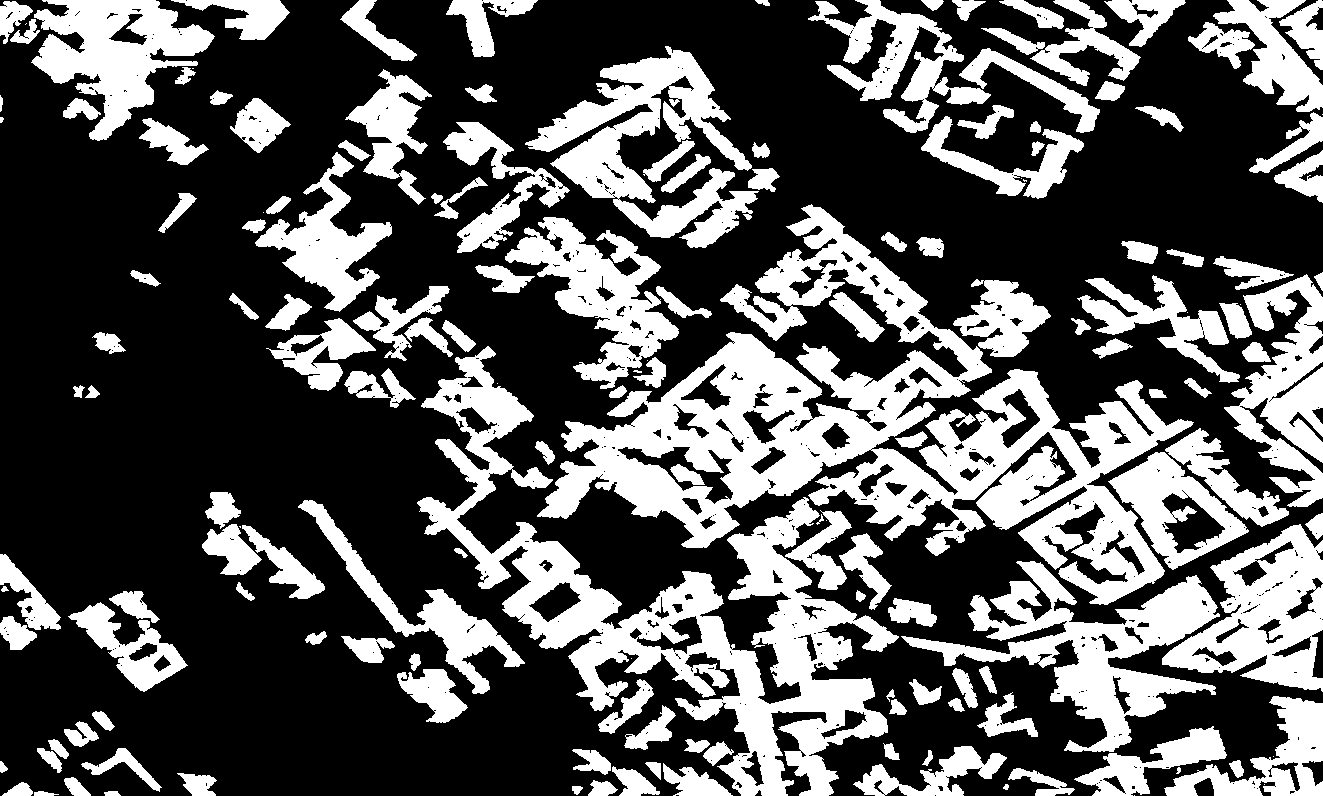}%
		}
	\caption{Common map of the produced building masks using the proposed annotation schemes and the reference ground truth GT map: (a) OSM-Ref $ \cap $ GT; (b) Opt-Ref $ \cap $ GT.}
	\label{fig:CommonOSMOpt}
\end{figure}

\begin{figure}[t]
	\centering
	\subfloat[]{\includegraphics[width=0.5\textwidth]{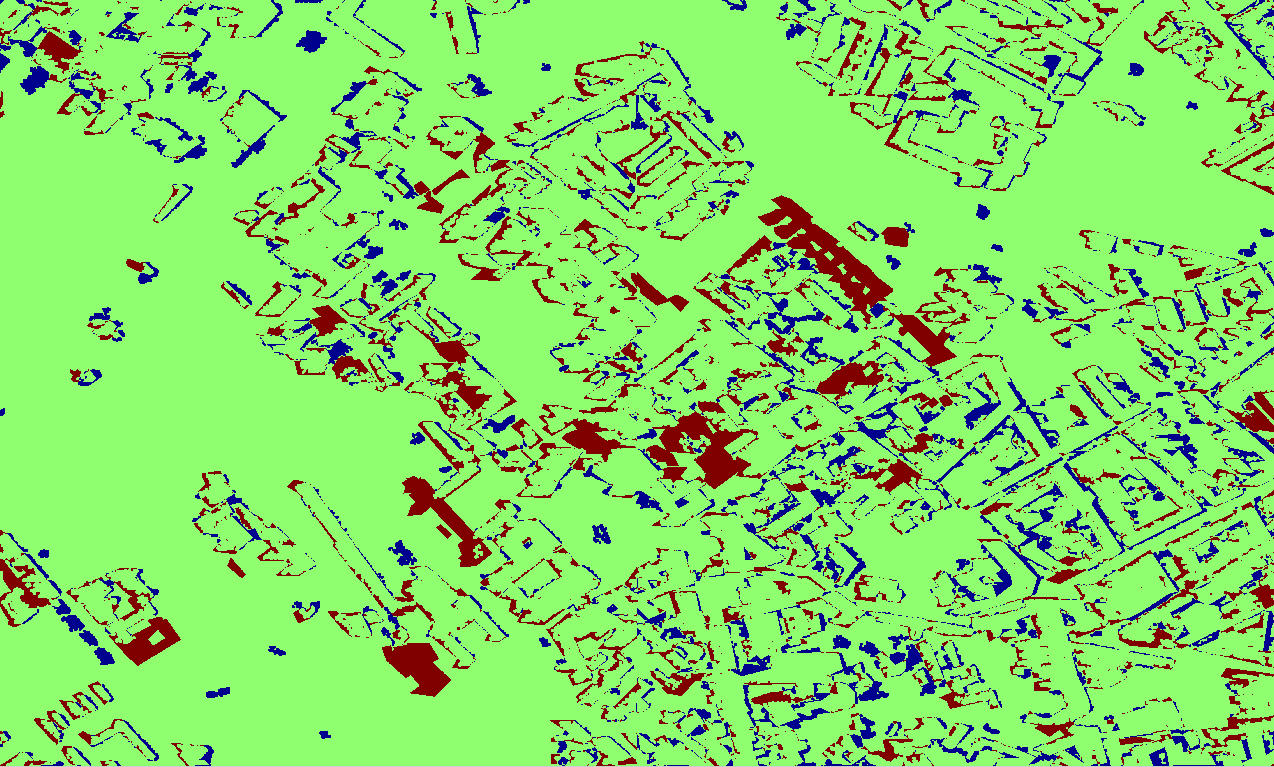}%
		}
	\hfil
	\subfloat[]{\includegraphics[width=0.5\textwidth]{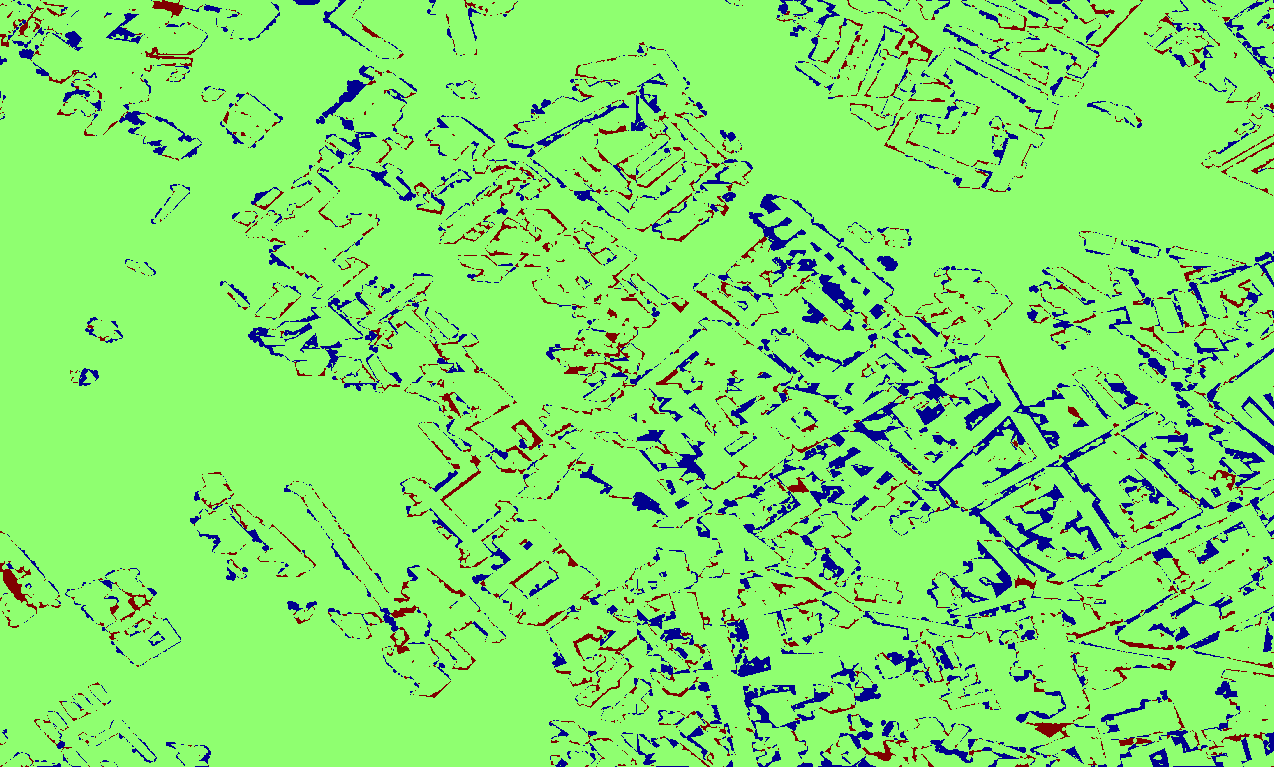}%
		}
	\caption{Difference map generated by subtracting the results of generated SAR annotations or training samples from the manually annotated ground truth GT mask: (a) OSM-Ref – GT; (b) Opt-Ref – GT. Note that the green pixels indicate no difference between the generated and ground truth masks whereas the red pixels indicate missing buildings while blue indicates the pixels labelled as belonging to buildings using the proposed annotated schemes but not present in the reference GT mask.}
	\label{fig:DiffOSMOpt}
\end{figure}

\begin{figure}[t]
	\centering
	\includegraphics[width=3.5in]{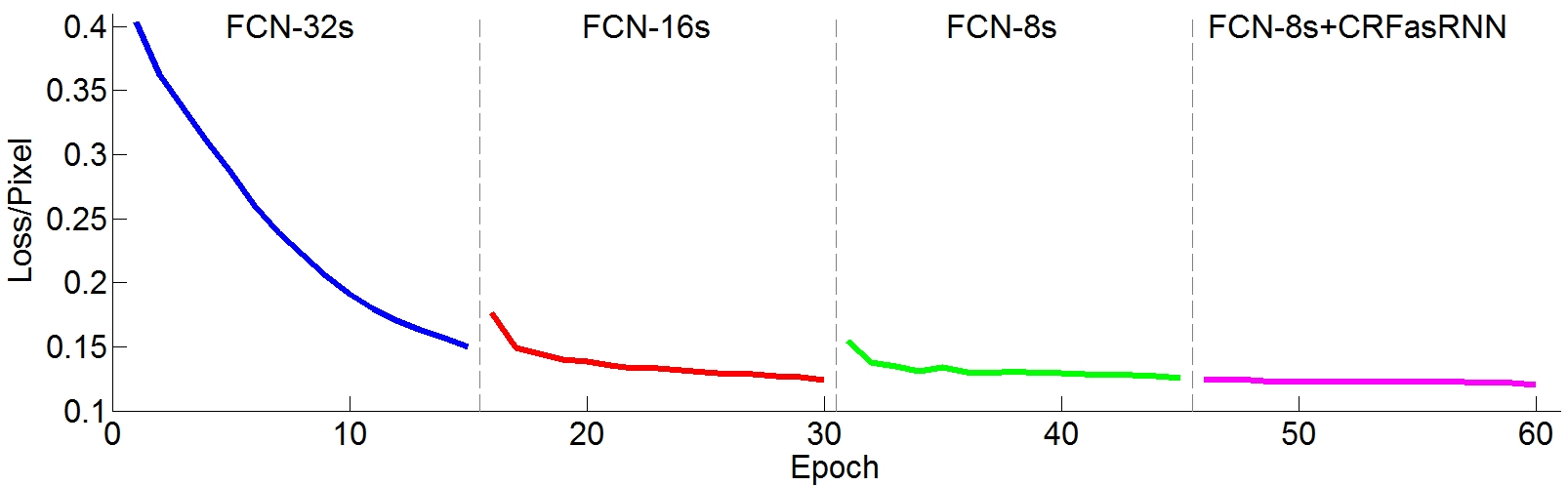}
	\caption{Learning curves across different stages of the network. The loss is normalized by dividing with the number of pixels of the training image. One Epoch represents all training images being passed through the network once.}
	\label{fig:LossCurves}
\end{figure}

\begin{figure*}[t]
	\centering
	\includegraphics[width=\textwidth]{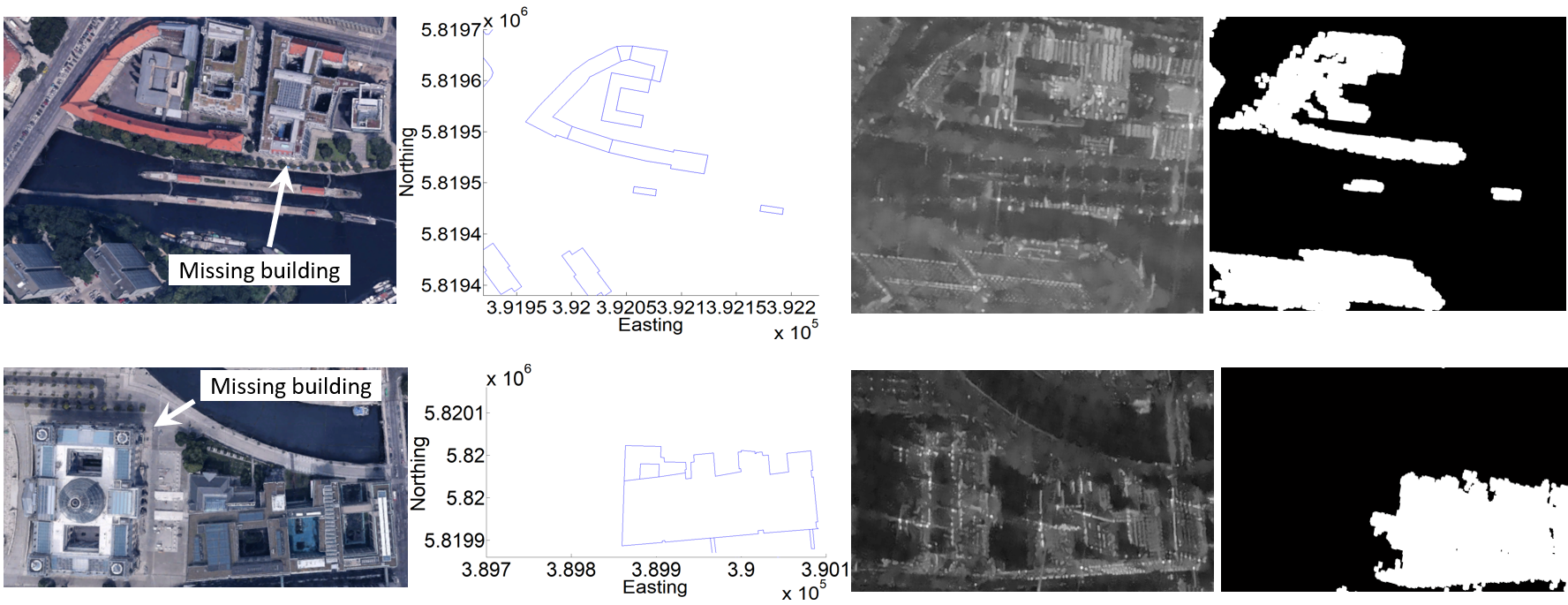}
	\caption{Missing buildings in the 2-D OSM GIS data. The 1st column shows the optical images of the buildings which are missing in the OSM polygonal data as shown in UTM coordinates in the 2nd column. The 3rd column presents the corresponding SAR images while the fourth column shows the ground truth generated by projecting the building points $ - $ extracted using auxiliary OSM GIS data $ - $ in the TomoSAR point clouds to the SAR image coordinates}
	\label{fig:Missing_Buildings}
\end{figure*}

\begin{figure*}[t]
	\centering
	\subfloat[]{\includegraphics[width=0.23\textwidth]{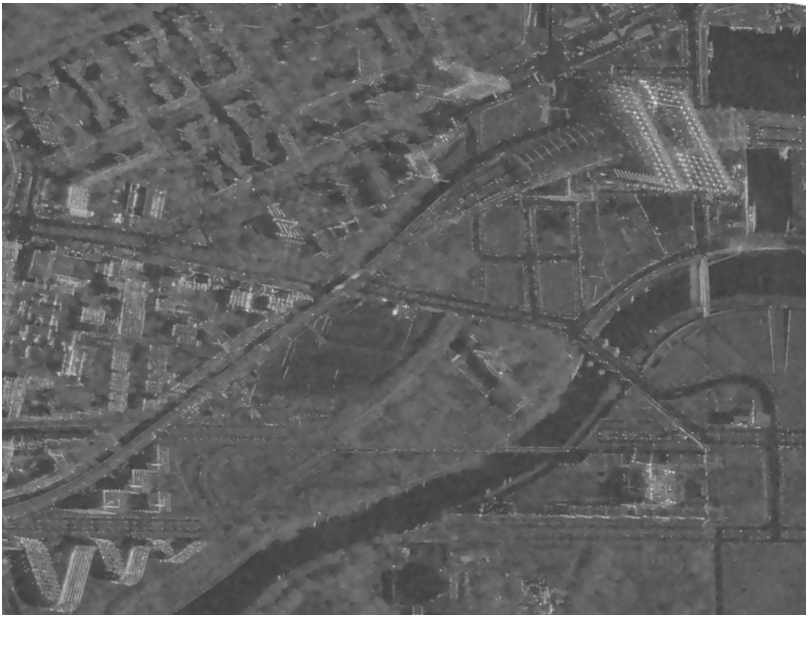}
		\label{fig:FalseRailway_a}}
	\hfil
	\subfloat[]{\includegraphics[width=0.25\textwidth]{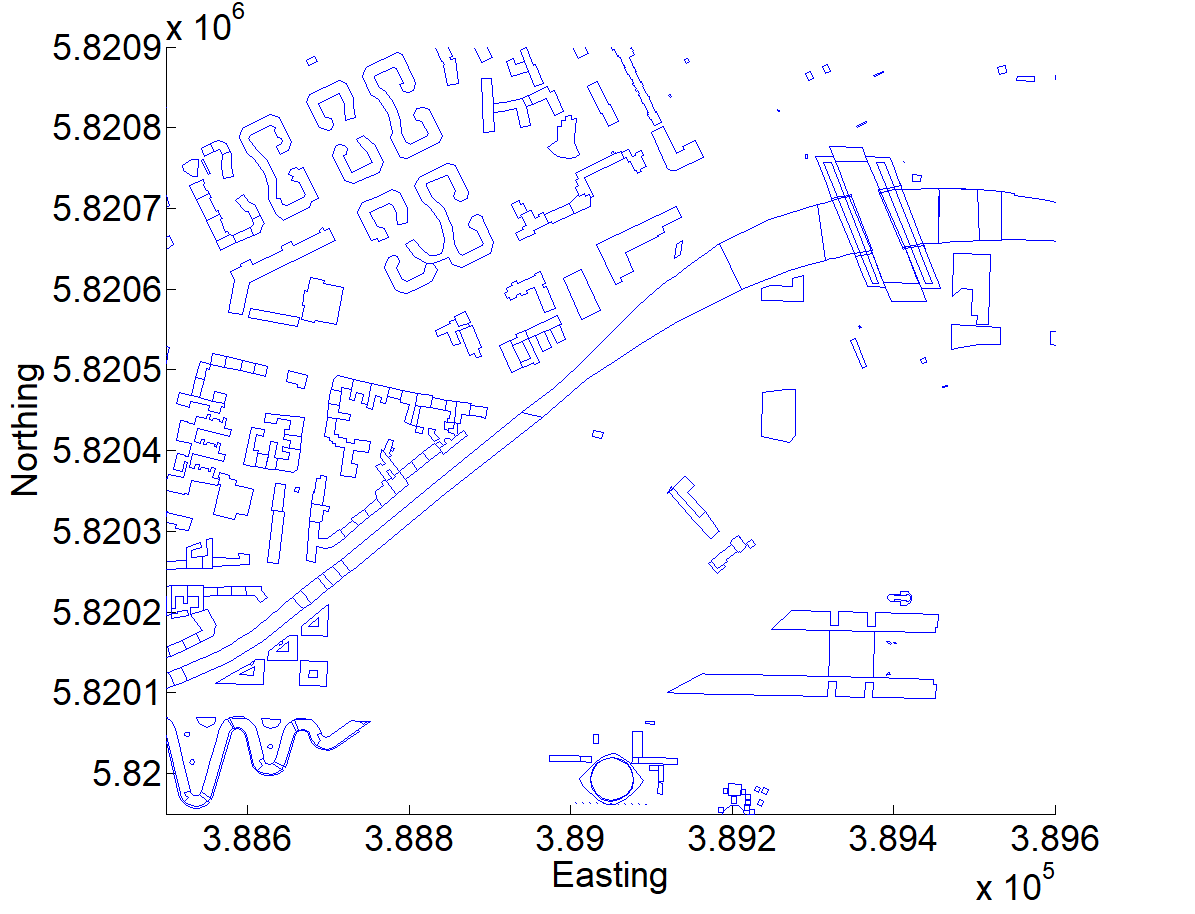}
		\label{fig:FalseRailway_b}}
	\hfil
	\subfloat[]{\includegraphics[width=0.22\textwidth]{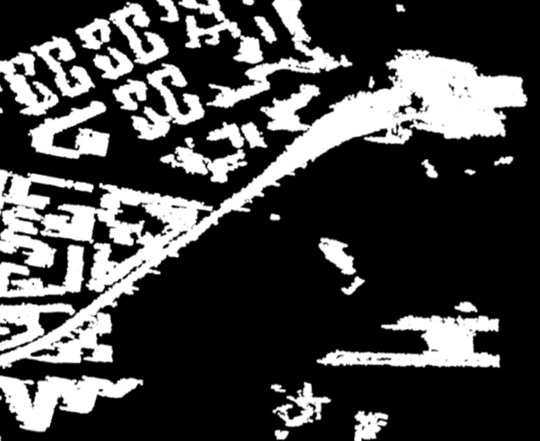}
		\label{fig:FalseRailway_c}}
	\hfil
	\subfloat[]{\includegraphics[width=0.225\textwidth]{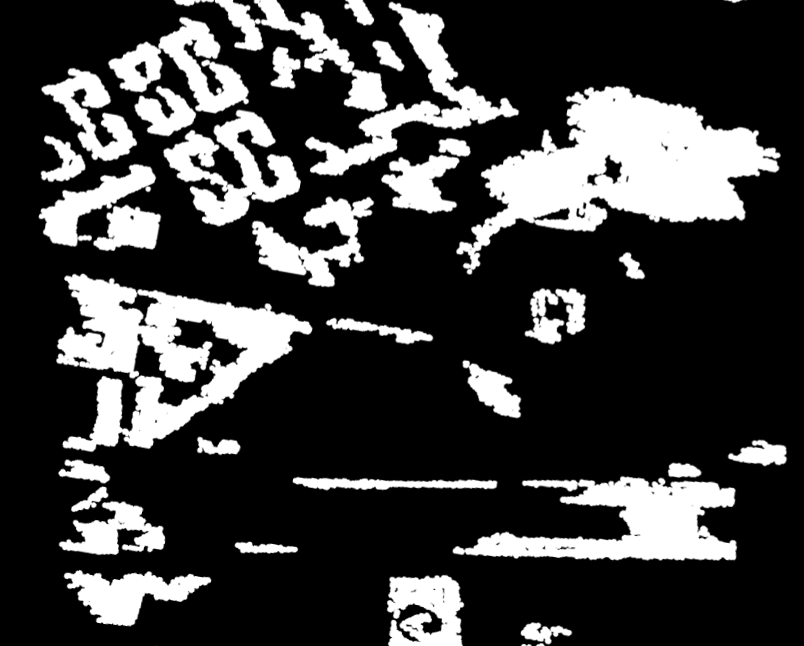}
		\label{fig:FalseRailway_d}}
	\caption{Close-up views of Berlin central station to show the false labeling in the ground truth generated using TomoSAR point cloud and the auxiliary OSM data: (a) presents the SAR image of the area of interest (Berlin central station); (b) shows the 2-D OSM building polygons. It can be seen that the railway track originating from the Berlin central station is falsely characterized as building structure in the OSM data. Consequently, TomoSAR points belonging to this track is misclassified as building points and when projected back to SAR image coordinates yields false labelling as depicted in (c); (d) provides a close-up view of the ground truth (labelled SAR image) of the same area generated by projecting the building points $ - $ extracted using optical image classification scheme \cite{wang_fusing_2017} $ - $ in the TomoSAR point clouds to the SAR image coordinates. In contrast, the railway track is now correctly labelled as non-building in the generated ground truth.}
	\label{fig:FalseRailway}
\end{figure*}

\subsection{Preparation of Training Data}\label{sec:PTD}
We prepared the dataset for training by taking 11 out of 16 of the labeled input images covering almost the whole city of Berlin (using OSM + TomoSAR point cloud) and created patches of $256 \times 256$ pixels with an overlap of 32 pixels. Further, these patches are augmented by flipping and rotation. Finally, we got 26,312 image patches for training and used the remaining 5 out of 16 of the labeled input images for testing. In case of Optical + TomoSAR point cloud, we vertically sliced the highlighted SAR image region depicted in Figure \ref{fig:MainImage42_Optical} in four equal parts and took the first and last for testing/validation and the two in the center for training. It is also important to mention that to reduce speckle effect, we first performed non-local filtering of the SAR images prior to training using the algorithm as proposed in \cite{baier_nonlocal_2016}.

\subsection{Performance Evaluation of the Trained Network}\label{sec:PE}

\subsubsection{Evaluation Metrics} \label{sec:EM}
To evaluate the performance of different networks, we use the metrics that are variations on pixel accuracy and region intersection over union (IU) and are commonly used for evaluating semantic segmentation and scene parsing algorithms \cite{long_fully_2015} \cite{shelhamer_fully_2016}. For each class, the Intersection over Union (IU) score is computed as: $
\frac{t_p}{t_p + f_p + f_n}$ where $ t_p $ (true positives) are the number of correctly classified pixels, $ f_p $ (false positives) are the number of wrongly classified pixels and $ f_n $ (false negatives) are the number of pixels wrongly not classified as belonging to a particular class. If we denote $ n_{ij} $ as the number of pixels of class $ i $ predicted to belong to class $ j $, $ n_N $ as the number of classes, and $ t_i $ as the total number of pixels belonging to class $ i $, then the following evaluation metrics have been computed \cite{long_fully_2015} \cite{shelhamer_fully_2016}

Pixel Accuracy (PA): $ \dfrac{\sum_i n_{ii}}{\sum_i t_i} $

Mean Accuracy (MA): $ \left( \dfrac{1}{n_N}\right) \sum_i \dfrac{n_{ii}}{t_i}  $ 	

Mean IU (MIU): $ \left( \dfrac{1}{n_N}\right) \sum_i \dfrac{n_{ii}}{t_i + \sum_j n_{ji} - n_{ii}} $

Frequency Weighted IU (FWIU): \begin{center}
	$ \left( \sum_{k} t_k \right)^{-1}  \sum_i \dfrac{t_i n_{ii}}{t_i + \sum_j n_{ji} - n_{ii}} $
\end{center}

In addition to the above four metrics, the following two metrics have also been computed:

False Alarm Rate (FAR): $ \dfrac{f_p}{t_p} \equiv \dfrac{\sum_i \sum_j n_{ij}}{\sum_i n_{ii}}$

Quality Rate (QR): \begin{center}
	$ \dfrac{t_p}{t_p + f_p + f_n} \equiv \dfrac{\sum_i n_{ii}}{\sum_i (t_i + \sum_j n_{ji} - n_{ii})} $
\end{center}

\subsubsection{Results Analysis} \label{sec:RA}

The experimental results have been obtained after applying \textit{staged} training where the results obtained after single-stream, then upgraded to two-stream and three-stream are depicted as FCN-32s (32$ \times $ upsampled prediction), FCN-16s (16$ \times $ upsampled prediction), and FCN-8s (8$ \times $ upsampled prediction) respectively. In each respective stage, the network is learned from end-to-end in a cascaded manner i.e., all initialization parameters of the previous stage are fed as input to the subsequent one. Let us denote the automatically generated annotated dataset using OSM + TomoSAR point cloud as OSM-Ref and using Optical classification + TomoSAR point cloud as Opt-Ref. Table \ref{table:OSM-Ref-only} and Table \ref{table:OSM-Ref-with-OSM-GT} depict the results acquired over the whole area of Berlin in different stages of the network architecture. In Table \ref{table:OSM-Ref-only}, for testing/validation, we analyzed the network performance by computing evaluation metrics over (untrained) 5/16 sub image patches annotated using OSM + TomoSAR point cloud (i.e., OSM-Ref). As mentioned earlier, the OSM dataset is prone to errors introduced as a consequent of crowd sourcing, therefore for fair evaluation of network architecture, we needed to prepare a more accurate annotated dataset (denoted as OSM-GT) for test sub images.

To prepare such a reference annotated dataset, we manually inserted missing buildings and removed parts of other structures e.g., railway tracks misclassified as buildings (see Figures \ref{fig:MainImage42} and \ref{fig:MainImage42_labelled}). Table \ref{table:OSM-Ref-with-OSM-GT} depicts evaluation results over untrained sub image patches using OSM-GT for testing/validation. For both Tables \ref{table:OSM-Ref-only} and \ref{table:OSM-Ref-with-OSM-GT}, we see the improvement in network performance in each subsequent stage. In general, the upgraded three-stream FCN-8s with CRF-RNN tends to show superior performance in distinguishing buildings from non-buildings. It is important to mention that one may argue here that since the OSM-Ref is used for training, the prediction should be more close to the OSM-Ref instead to OSM-GT. The other way around reason is merely due to the fact that the trained CNN architecture correctly recognizes the missing buildings and was able to differentiate the railway tracks from buildings mainly because the training samples contain less portions of the railway tracks which were wrongly classified as buildings in the OSM data. 

Similarly, Table \ref{table:Opt-Ref-FCN-8s-CNN-RNN} shows the evaluation results with the FCN-8s with CRF-RNN network architecture with two annotated test images OSM-GT and Opt-Ref where the later is result of automatically annotated test sub images generated using Optical classification + TomoSAR point cloud. These quantitative results are obtained using the training sample division as reported in section \ref{sec:PTD}.

\begin{table}[t]
	\caption{Accuracy analysis of obtained results using different stages of the trained network with following details: Training \& testing/validation using OSM-Ref data.} 
	\centering  
	\begin{tabularx}{3.5in}{c c c c c c c} 
		\hline\hline                        
		Network Architecture & PA & MA & MIU & FWIU & FAR & QR \\ [0.5ex] 
		\hline                  
		FCN-32s & 83.28 & 82.40 & 69.47 & 71.98 & 20.39 & 71.58  \\  
		FCN-16s & 83.46 & 82.43 & 69.70 & 72.22 & 20.12 & 71.85 \\
		FCN-8s & 83.49 & 82.45 & 69.75 & 72.27 & 20.06 & 71.90 \\
		FCN-8s (CRF-RNN) & 83.54 & 82.60 & 69.86 & 72.35 & 20.00 & 71.97 \\ [1ex]    
		\hline  
	\end{tabularx}
	\label{table:OSM-Ref-only} 
\end{table}

\begin{table}[t]
	\caption{Accuracy analysis of obtained results using different stages of the trained network with following details: Training using OSM-Ref dataset \& testing/validation using OSM-GT.} 
	\centering  
	\begin{tabularx}{3.5in}{c c c c c c c} 
		\hline\hline                        
		Network Architecture & PA & MA & MIU & FWIU & FAR & QR \\ [0.5ex] 
		\hline                  
		FCN-32s & 89.87 & 91.32 & 79.89  & 82.16  & 11.36  & 81.72  \\  
		FCN-16s & 91.35 & 92.78 & 82.52  & 84.54  & 9.54  & 84.17  \\
		FCN-8s & 91.52 & 92.97  & 82.81  & 84.81  & 9.34  & 84.45  \\
		FCN-8s (CRF-RNN) & 92.13 & 93.84 & 83.97 & 85.82  & 8.61  & 85.48  \\ [1ex]    
		\hline  
	\end{tabularx}
	\label{table:OSM-Ref-with-OSM-GT} 
\end{table}

\begin{table}[t]
	\caption{Accuracy analysis of obtained results using FCN-8s with CRF-RNN network architecture utilizing automatically generated annotated data using Optical classification and TomoSAR point cloud, denoted as Opt-Ref, as training data and OSM-GT and Opt-Ref as testing/validation data.} 
	\centering  
	\begin{tabularx}{3.5in}{c c c c c c c c} 
		\hline\hline                        
		Training & Testing & PA & MA & MIU & FWIU & FAR & QR \\ [0.5ex] 
		\hline                  
		Opt-Ref & OSM-GT & 78.35 & 69.30 & 56.97 & 63.74 & 28.49 & 64.94  \\  
		Opt-Ref & Opt-Ref & 83.23 & 79.57 & 66.45 & 72.45 & 20.55 & 71.57 \\ [1ex]    
		\hline  
	\end{tabularx}
	\label{table:Opt-Ref-FCN-8s-CNN-RNN} 
\end{table}

\begin{figure}[t]
	\centering
	\includegraphics[width=3.5in]{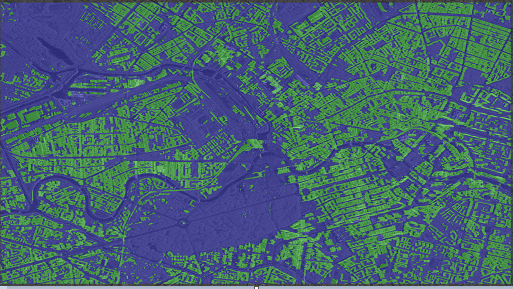}
	\caption{Input SAR image of Berlin city as depicted in Figure \ref{fig:MainImage42} with overlay of the semantic segmentation. Results computed using OSM-Ref annotated dataset with FCN-8s with CRF-RNN network.}
	\label{fig:MainImage42_Overlay_OSM_low}
\end{figure}

\begin{figure}[t]
	\centering
	\includegraphics[width=3.5in]{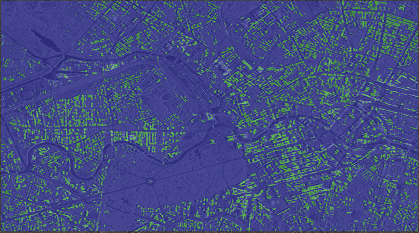}
	\caption{Input SAR image of Berlin city as depicted in Figure \ref{fig:MainImage42} with overlay of the semantic segmentation. Results computed using Opt-Ref annotated dataset with FCN-8s with CRF-RNN network.}
	\label{fig:MainImage42_Overlayed_Optical_low}
\end{figure}

\begin{figure*}[t]
	\centering
	\includegraphics[width=\textwidth]{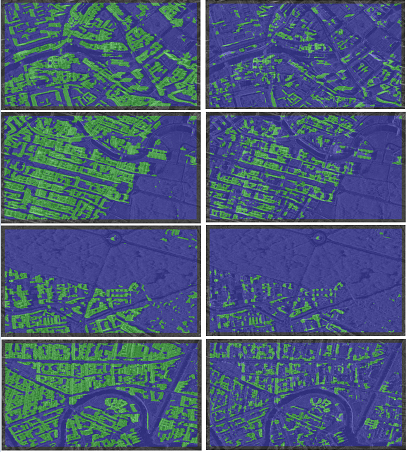}
	\caption{Close-up views of Figure \ref{fig:MainImage42_Overlay_OSM_low} (1st column) and Figure \ref{fig:MainImage42_Overlayed_Optical_low} (2nd column). The 1st column depicts the input image with overlay of the semantic segmentation result. The results have been computed using OSM-Ref annotated dataset with FCN-8s with CRF-RNN network over different test sub image patches. The 2nd column depicts the input image with overlay of the semantic segmentation results. The results have been computed using Opt-Ref annotated dataset with FCN-8s with CRF-RNN network over different test sub image patches.}
	\label{fig:CloseUpOverlays}
\end{figure*}

\begin{figure*}[t]
	\centering
	\includegraphics[width=\textwidth]{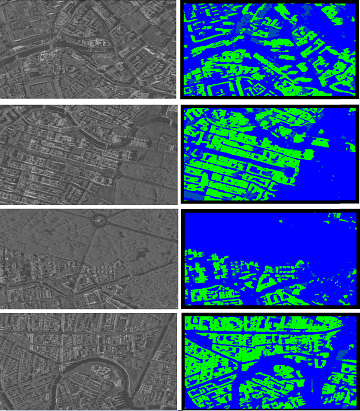}
	\caption{Results of semantic segmentation computed using OSM-Ref annotated dataset with FCN-8s with CRF-RNN network over different test sub image patches. The first column shows the different SAR test sub image patches while the 2nd column depicts the difference image of the semantic segmentation result and the manually corrected ground truth (Training: OSM-Ref; Testing: OSM-GT; Network FCN-8s with CRF-RNN). The light green region in the difference map corresponds to true positives while the dark green regions are false negatives (negligible here). Light red, on the other hand, are the true negatives while dark red regions corresponds false positives.}
	\label{fig:DifferenceImageOSM}
\end{figure*}

\begin{figure*}[t]
	\centering
	\includegraphics[width=\textwidth]{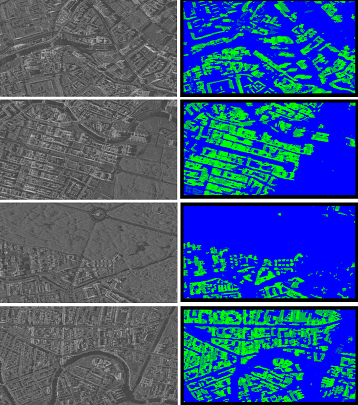}
	\caption{Results of semantic segmentation computed using Opt-Ref annotated dataset with FCN-8s with CRF-RNN network over different test sub image patches. The first column shows the different SAR test sub image patches while the 2nd column depicts the difference image of the semantic segmentation result and the manually corrected ground truth (Training: Opt-Ref; Testing: OSM-GT; Network FCN-8s with CRF-RNN). The light green region in the difference map corresponds to true positives while the dark green regions are false negatives. Light red, on the other hand, are the true negatives while dark red regions corresponds false positives.}
	\label{fig:DifferenceImageOpt}
\end{figure*}

Figures \ref{fig:MainImage42_Overlay_OSM_low} and \ref{fig:MainImage42_Overlayed_Optical_low} shows the result of FCN-8s with CRF-RNN trained using OSM-Ref and Opt-Ref respectively overlaid onto the SAR image of Figure \ref{fig:MainImage42} covering almost the whole region of Berlin. Again, these qualitative results are obtained using the training sample division as reported in section \ref{sec:PTD}. Figure \ref{fig:CloseUpOverlays} shows the close-up results over different test sub image patches while Figures \ref{fig:DifferenceImageOSM} and \ref{fig:DifferenceImageOpt} shows the corresponding difference maps. The light green region in the difference map corresponds to common regions i.e., true positives while the dark green regions are buildings that have not been detected by the network i.e., false negatives. Light blue, on the other hand, are the true negatives while dark blue regions corresponds to wrongly classified buildings i.e., false positives. With OSM-Ref trained network, we hardly see any dark green regions implicitly implying high degree of completeness (see Figure \ref{fig:DifferenceImageOSM}). In contrast, for Opt-Ref trained network we have a fair amount of dark green regions depicting miss detections (see Figure \ref{fig:DifferenceImageOpt}). The main reason for this is that the network has been trained with less number of training samples in case of Opt-Ref as compared to OSM-Ref (see section \ref{sec:PTD}). 

Nevertheless, for comparison and to provide accurate and fair accuracy analysis, we also trained the network separately using both OSM-Ref and Opt-Ref annotated SAR building masks in a controlled manner. We carefully designed the experiment by using the same (geographic) region, network parameters, and the size of the training patches. For evaluation, we used the manually prepared ground truth testing mask GT depicted in Figure \ref{fig:GT}. Table \ref{table:OSM-Opt-Analysis} shows the evaluation results obtained by training the FCN-8s with CRF-RNN network architecture with both OSM-Ref and Opt-Ref annotated masks and tested using GT. The quantitative analysis of this experiment demonstrates that the network trained using both the annotation masks reveals similar performance. However, since there are completeness issues with OSM data, the Opt-Ref annotation is slightly better. In spite of this, in comparison to Opt-Ref, the generation of OSM-Ref annotation mask is much easier to obtain and has the potential to produce the large-scale SAR annotation masks. The accuracy of such kind of masks can however be improved by replacing the OSM data with more accurate cadastral data (2-D footprints), if available from other sources, e.g., city administration etc. 

\begin{table}[t]
	\caption{Accuracy analysis of obtained results using FCN-8s with CRF-RNN network architecture utilizing OSM-Ref and Opt-Ref annotated masks. The network parameters such as learning rate, momentum, weight decay etc. are same and provided in the beginning of section \ref{sec:ATA}. GT corresponds to the manually prepared testing ground truth mask depicted in Figure \ref{fig:GT}.} 
	\centering  
	\begin{tabularx}{3.5in}{c c c c c c c c} 
		\hline\hline                        
		Training & Testing & PA & MA & MIU & FWIU & FAR & QR \\ [0.5ex] 
		\hline                  
		OSM-Ref & GT & 82.80 & 81.25 & 69.06 & 70.82 & 20.77 & 70.65  \\  
		Opt-Ref & GT & 83 & 81.58 & 69.42 & 71.13 & 20.45 & 70.95 \\ [1ex]    
		\hline  
	\end{tabularx}
	\label{table:OSM-Opt-Analysis} 
\end{table}

\subsubsection{Analysis of the network}
Figure \ref{fig:LossCurves} demonstrates the learning curves across different stages of the network using ReLU activation function. As shown in the figure, we can make use of a fairly high learning rate to train the staged network for detecting building regions without the risk of divergence. The jumps between different stages of the network architecture originates by the fact of having a new part at the end of the network that is just initialized but not trained at all (due to staged training).

\subsubsection{Hardware and processing time}
All the experiments have been conducted on a GPU equipped personal computer with following details: Intel Core i7 @ 3.7GHz and 32GB RAM. For one test image of the dimension 20626 x 11472 covering almost the whole area of Berlin, it took in average around 259 seconds. Such an evident fast speed of deep learning architectures is very important in practical scenarios. Also, the downside of deep learning architectures (i.e., long training times) is becoming increasingly ignorable with rapid development in the hardware technology particularly in GPUs.

\section{Discussion} \label{sec:Discussion}
The experiments presented in this paper show a variety of things:

\begin{itemize}
	\item It demonstrated that it is possible to automatically generate reference datasets with the potential to be produced globally opening new perspectives of producing benchmark SAR reference datasets. Other method of choice to generate such a reference dataset may be obtained by exploiting simulation based methods as proposed e.g., in \cite{auer_ray-tracing_2010} \cite{tao_automatic_2014}. However such methods have their own limitations in a sense, they typically require accurate models (\mbox{3-D} building models and/or accurate digital surface models) to precisely generate such ground truth data which, in most cases, is not available.
	\item Deep learning architectures are greedy in terms of training data limiting their potential application. However, with the possibility of producing large-scale annotated datasets, application of different deep learning network architectures are possible for classification of built-up areas in SAR images.
	\item In the case of OSM data, although the completeness and correctness of the OSM data is fairly good but has not yet reached to a level where it covers the whole globe. Nevertheless, at least in the developed countries such a data has the potential to be used either directly as the reference/ground truth dataset or to generate training data (i.e., the labeled buildings masks as demonstrated in our case) where it is difficult to obtain such information with other interactive/expert methods.
	\item It is also worth to mention that both the automatic annotation results are produced using TomoSAR point clouds. Due to complex multiple scattering and different microwave scattering properties of the objects in the scene which possess different geometrical and material features, TomoSAR point clouds exhibit some special characteristics such as low positioning accuracy, high number of outliers, gaps in the data and rich façade information due to the side looking geometry. These properties makes classification of TomoSAR point clouds a challenging task. With the aid of additional auxiliary information, the problem is rectified.
	\item Hypothetically, the automatic annotation results could be improved with higher density of TomoSAR points because when projected to SAR coordinates (azimuth and range), a denser map of the reference would be generated. TomoSAR point density is however dependent on several factors e.g., the geometrical properties of objects appearing in the scene, number of SAR images used for tomographic reconstruction etc. In current scenario, the effect of low point density is reduced by densifying the resulting building mask using mathematical image dilation operation.
	\item Lastly, the capability to produce automatic large area annotations together with their exploitation to detect buildings in SAR imagery may benefit the field of SAR based (e.g., D-InSAR or TomoSAR) risk management against potential threats (including subsidence, landslides etc.) by performing building damage/vulnerability analysis e.g., as depicted in \cite{cascini_detection_2013} \cite{peduto_empirical_2017}.    
\end{itemize}

\section{Conclusion \& Outlook}
In this article, we have presented a deep learning based network architecture that is able to classify buildings from non-buildings in SAR images. Two automated annotation methods able to generate reference building masks for training and testing the classifier have been presented. The methods of automated annotation are generic and have the potential towards generation of large scale SAR reference datasets. The annotated building masks have been utilized to construct and train the deep fully Convolution Neural Networks with an additional Conditional Random Field represented as a Recurrent Neural Network to detect building regions in a single (non-locally filtered) SAR image with mean accuracy of around 93.84\%. The presented results are expected to further stimulate the research interest in exploiting SAR imagery using deep learning network architectures.

The results of this study are promising but still there are things which could be addressed in future. For instance, the heights of individual buildings could be retrieved/estimated by identifying layover regions in the obtained CNN based detection results. One application of such estimation is in reducing the number of images required for accurate tomographic reconstruction as demonstrated in \cite{zhu_joint_2015}. Also, in the current study, we aimed at detecting buildings for which we utilized/generated OSM based annotated building masks for training and testing/validation. In future, such annotated masks could also be produced using other objects in the OSM dataset e.g., roads, coastlines etc.


%



\section*{Acknowledgment}
The authors would like to thank Gerald Baier from German Aerospace Center (DLR), Germany to perform non local filtering of the SAR image of Berlin used in this article.This work is supported by the European Research Council (ERC) under the European Union’s Horizon 2020 research and innovation programme (grant agreement No [ERC-2016-StG-714087], Acronym: \textit{So2Sat}) and Helmholtz Association under the framework of the Young Investigators Group ``SiPEO'' (VH-NG-1018, www.sipeo.bgu.tum.de). The authors gratefully acknowledge the Gauss Centre for Supercomputing e.V for providing computing time at the GCS Supercomputer SuperMUC at the Leibniz Supercomputing Centre (Project ID: pr53ya). \cite{Xu2017} \cite{brunner_building_2010}

\ifCLASSOPTIONcaptionsoff
  \newpage
\fi



%

	
	

\bibliographystyle{IEEEtran}
\bibliography{SAR_CNN_Updated}

%








\end{document}